\documentclass[twocolumn,times,trackchanges]{aastex62}
\hypersetup{linkcolor=red,citecolor=blue,filecolor=cyan,urlcolor=magenta}
\usepackage{xspace}
\usepackage{multirow}

\newcommand{\FeK}{\mbox{Fe K$\alpha$}\xspace}
\newcommand{\fluxu}{\mbox{ergs cm$^{-2}$ s$^{-1}$}\xspace}
\newcommand{\ngcfour}{\mbox{NGC 4151}\xspace}
\newcommand{\xmm}{\mbox{XMM-Newton}\xspace}
\newcommand{\nustar}{\mbox{NuSTAR}\xspace}
\newcommand{\suzaku}{\mbox{Suzaku}\xspace}
\newcommand{\chandra}{\mbox{Chandra}\xspace}
\newcommand{\xsm}[2]{{\tt #1$_{\rm #2}$}\xspace}
\newcommand{\rg}{$r_g$\xspace}
\newcommand{\gone}{${g_{1}}$\xspace}

\defcitealias{2012MNRAS.422..129Z}{Z12}

\graphicspath{{./}{figures/}}

\received{January 1, 2018}
\revised{January 7, 2018}
\accepted{\today}

\shorttitle{Revisiting NGC 4151}
\shortauthors{Zoghbi et al.}

\begin{document}

\title{Revisiting The Spectral and Timing Properties of NGC 4151 }

\email{abzoghbi@umich.edu}

\author[0000-0002-0572-9613]{A. Zoghbi}
\affil{Department of Astronomy, University of Michigan, Ann Arbor, MI 48109, USA}

\author{J. M. Miller}
\affiliation{Department of Astronomy, University of Michigan, Ann Arbor, MI 48109, USA}

\author{E. Cackett}
\affiliation{Department of Physics, Wayne State University, Detroit, MI 48201, USA}

\begin{abstract}
NGC 4151 is the brightest Seyfert 1 nucleus in X-rays. It was the first object to show short time delays in the Fe K band, which were attributed to relativistic reverberation, providing a new tool for probing regions at the black hole scale. Here, we report the results of a large XMM-Newton campaign in 2015 to study these short delays further. Analyzing high quality data that span time scales between hours and decades, we find that neutral and ionized absorption contribute significantly to the spectral shape. Accounting for their effects, we find no evidence for a relativistic reflection component, contrary to early work. Energy-dependent lags are significantly measured in the new data, but with an energy profile that does not resemble a broad iron line, in contrast to the old data. The complex lag-energy spectra, along with the lack of strong evidence for a relativistic spectral component, suggest that the energy-dependent lags are produced by absorption effects. The long term spectral variations provide new details on the variability of the narrow \FeK line . We find that its variations are correlated with, and delayed with respect to, the primary X-ray continuum. We measure a delay of $\tau = 3.3^{+1.8}_{-0.7}$ days, implying an origin in the inner broad line region (BLR). The delay is half the H$\beta$ line delay, suggesting a geometry that differs slightly from the optical BLR.
\end{abstract}


\section{Introduction} \label{sec:intro}
The X-ray emission in the active nuclei of Seyfert galaxies is thought to be produced in a corona that Compton scatters optical/UV photons from the accretion disk \citep{1993ApJ...413..507H} or in the base of a jet \citep{2005ApJ...635.1203M} close to the central supermassive black hole. The radiation can be seen directly, through absorbing systems, or when it is reflected and scattered in surrounding media, leading to a diversity in spectral shapes and variability behavior \citep[e.g.][]{2014ApJ...788...76W,2015ApJ...804..107R,2018MNRAS.473.4377W}.

Absorption by gas with column densities in the levels $10^{21}-10^{24}$ $\rm{cm}^{-2}$ is often observed, with ionization that range from neutral to partially and fully ionized gas \citep[warm absorber][]{1984ApJ...281...90H,1997MNRAS.286..513R}. Multiple layers of absorption are often observed \citep[e.g.][]{2016A&A...587A.129E}. Additionally, several occultation events associated with Compton thin and thick clouds have been observed \citep{2007ApJ...659L.111R, 2013MNRAS.436.1588S,2014MNRAS.439.1403M}.

On the emission side, iron K$\alpha$ emission lines (\FeK) at 6.4 keV are observed ubiquitously in AGN \citep{1994MNRAS.268..405N,2001ApJ...550..261W,2009ApJ...690.1322W,2010ApJS..187..581S}. They are produced when the central X-ray source illuminates cold and dense material in its surroundings, making it a powerful probe of the circumnuclear environment in AGN. 

The line is produced in optically thick material outside the direct line of sight. It is generally accompanied by a Compton reflection continuum \citep{1988MNRAS.233..475G,1988ApJ...335...57L,1991MNRAS.249..352G}, and its origin in this case is likely the outer disk or the torus \citep{1994ApJ...420L..57K,2004A&A...422...65B,2006MNRAS.368L..62N}. In some cases, the line appears to be emitted from a Compton thin gas, likely the Broad Line Region (BLR) \citep[e.g.][]{2008MNRAS.389L..52B}

The narrow core of the line is often accompanied by an additional, relativistically broadened, component that originates much closer to the black hole \citep{1989MNRAS.238..729F,2003PhR...377..389R,2007MNRAS.382..194N,2007ARA&A..45..441M,2015ApJ...806..149K}. Measurement of the shape of the broad iron line and the reflection spectrum provides direct measurement of the black hole spin. The discovery of reverberation of the broad iron line opened a new exploration route \citep{2012MNRAS.422..129Z, 2016MNRAS.462..511K}. Short time delays between energy bands dominated by the iron line and the primary continuum are observed, suggesting that the the emission region is compact in size \citep[$<10$ \rg;][]{2014MNRAS.438.2980C}.

\ngcfour, due to its brightness, was the first object to show a putative relativistic reverberation signature using \xmm data \citep[][hereafter Z12]{2012MNRAS.422..129Z}. A relativistic reflection component was also inferred in the spectra (\citetalias{2012MNRAS.422..129Z}; \citealt{2015ApJ...806..149K, 2017A&A...603A..50B}). \ngcfour had one of the cleanest measurements of reverberation in AGN, so it was targeted with a large \xmm observing campaign in order to explore the nature of the lag. Here, we report the results from that campaign and from archival \xmm, \suzaku and \nustar observations, and in light of the new data, we revisit the spectral modeling and the lag interpretation.

\ngcfour (z = 0.00335) is one of the nearest archetypal Seyfert 1 galaxies. It is classified as a Seyfert 1.5 and it hosts one of the brightest AGN \citep{2000A&ARv..10..135U}. Its brightness in X-rays has made it a good target of all major X-ray telescopes \citep{1976ApJ...207L.159I,1989MNRAS.236..153Y,1992A&A...256L..38J,1995MNRAS.273..923P,1995ApJ...453L..81Y,1999NuPhS..69..481P,2000ApJ...545L..81O,2001ApJ...563..124Y,2003MNRAS.345..423S,2005ApJ...633..693K,2010MNRAS.408.1851L,2015ApJ...806..149K,2016ApJ...833..191C}. The X-ray spectral complexity of the nuclear emission in \ngcfour was apparent from the early observations \citep{1980ApJ...241L..13H,1989MNRAS.236..153Y}, while emission from extended regions was known to have a significant contribution below 1 keV \citep{1983ApJ...268..105E,1986MNRAS.218..685P}. The spectra of \ngcfour at energies above 10 keV have also been a testing laboratory for many coronal emission models \citep{1996MNRAS.283..193Z,1992ApJ...398L..37C,2010MNRAS.408.1851L}.

A description of the data used in this analysis is presented in section \ref{sec:data}. Spectral and timing analyses are presented in sections \ref{sec:spec} and \ref{sec:timing}, respectively. Section \ref{sec:narrow_lag} in particular presents the first direct measurement of the time delay between the {\em narrow} \FeK and the X-ray continuum. The results are contextualized and summarized in section \ref{sec:discuss}.

\section{Observations \& Data Reduction} \label{sec:data} \footnote{Codes and detailed analysis procedures are available at \url{https://zoghbi-a.github.io/Revisiting-NGC-4151-Data}.}
\xmm observed NGC 4151 several times over the last two decades. We reduced all the available 29 observations. Five of these had no useful science exposures or were dominated by background flares, leaving 24 observations. The most recent 9 exposures were taken in November and December 2015. The \xmm EPIC data is reduced using \texttt{epchain} and \texttt{emchain} in \textsc{SAS}. Multiple exposures within a single observations are combined. 

Source photons are extracted from circular regions of 50\arcsec\,radius centered on the source. Given the brightness of the source, photon pileup needs to be considered. We use the \texttt{epatplot} tool to check for the effect. The fraction of single and double events are calculated between 2--10 keV, the band relevant for the science in this work. If pileup is detected, the source photons are extracted from an annular region with an outer radius of 50\arcsec\,and an inner radius that we increase in steps of 0.5\arcsec\,starting from 3\arcsec,\ until the ratio of observed to predicted single and double photon events are within $3\sigma$ of unity. Background photons are extracted from circular regions similar to and away from the source. The RGS data, which we examine briefly, is reduced using \texttt{rgsproc}, the first and second order spectra are combined using \texttt{rgscombine} and grouped according to the optimal binning suggested by \cite{2016A&A...587A.151K}. All other EPIC spectra are grouped so that the detector resolution is over-sampled by a factor of 3, ensuring the minimum signal to noise ratio per bin is 6.
Details of the \xmm observations are shown in Table \ref{tab:obs_log}. For the spectral analysis presented in the following sections, we consider only observations with a minimum net exposure of 5 ks, hence 0657840201 and 0679780301 are not considered. The final set therefore includes 22 spectra.

\begin{deluxetable}{ccllll}
\tablecaption{Observation log. \xmm, \suzaku and \nustar data are separated by horizontal lines. R$_{\rm i}$ in the case of the \xmm data is the radius of circular region excluded when extracting the source photons to reduce photon pileup. The gain column gives the parameters of the \texttt{gain} model in \textsc{xspec}. The * symbol indicates that the exposure is short and not considered. $(^t)$ identifies the \xmm observations used in the fast timing analysis.\label{tab:obs_log}}
\tablehead{\colhead{\#} & \colhead{ObsID} & \colhead{R$_{\rm i}$} & \colhead{Gain} & \colhead{Exp.} & \colhead{Time} \\[-0.2cm]
\colhead{} & \colhead{} & \colhead{(\arcsec)} & \colhead{slope,offset} & \colhead{(ks)} & \colhead{(MJD)}}
\startdata
1$^t$   & 0112310101 & 0   & 1.0,   -0.004      & 21.0	& 51899.9 \\
2$^t$   & 0112830201 & 0   & 1.0,    0.00309    & 50.9	& 51900.8 \\
3$^t$   & 0112830501 & 0   & 1.001,  0.00106    & 17.6	& 51900.3 \\
4$^t$   & 0143500101 & 0   & 1.002, -0.00717    & 11.1	& 52784.2 \\
5$^t$   & 0143500201 & 3   & 1.003, -0.0112     & 12.7	& 52786.0 \\
6$^t$   & 0143500301 & 0   & 1.005, -0.00936    & 12.7	& 52786.8 \\
7$^t$   & 0402660101 & 0   & 1.002, -0.00202    & 28.0	& 53871.5 \\
8$^t$   & 0402660201 & 4   & 1.005, -0.0107     & 22.9	& 54069.0 \\
9	    & 0657840201 & 0   & *					& 2.2	& 55724.7 \\
10	    & 0657840301 & 3   & 1.005, -0.0138     & 5.6	& 55890.2 \\
11      & 0657840401 & 0   & 1.006, -0.00887	& 6.6	& 55904.2 \\
12	    & 0679780101 & 0   & 1.004, -0.00312    & 6.3	& 56060.3 \\
13	    & 0679780201 & 3   & 1.003, -0.00546    & 8.8	& 56088.2 \\
14	    & 0679780301 & 0   & *					& 3.8	& 56245.8 \\
15	    & 0679780401 & 0   & 1.007, -0.0123	    & 6.6	& 56271.7 \\
16$^t$  & 0761670101 & 3   & 1.009, -0.00496 	& 22.8	& 57338.9 \\
17$^t$  & 0761670201 & 3   & 1.011, -0.00643	& 24.0	& 57340.8 \\
18$^t$  & 0761670301 & 3   & 1.009, -0.00833	& 30.6	& 57342.8 \\
19$^t$  & 0761670401 & 3   & 1.010,   0.00486	& 21.9	& 57356.8 \\
20$^t$  & 0761670501 & 3   & 1.011, -0.0130	    & 26.9	& 57346.8 \\
21$^t$  & 0761670601 & 3   & 1.010, -0.00703	& 29.9	& 57348.8 \\
22$^t$  & 0761670701 & 0   & 1.010, -0.00984	& 29.4	& 57372.7 \\
23$^t$  & 0761670801 & 3   & 1.009, -0.00525 	& 29.7	& 57374.7 \\
24$^t$  & 0761670901 & 0   & 1.009, -0.00933	& 30.4	& 57378.7 \\ 
\hline
s1      & 701034010  & -   & -                  & 125.0 & 54087.8 \\
s2      & 707024010  & -   & -                  & 150.3 & 56242.9 \\
s3      & 906006010  & -   & -                  & 61.7  & 55882.7 \\
s4      & 906006020  & -   & -                  & 60.6  & 55913.7 \\
\hline
n1      & 60001111002 & -  & -                  & 21.9  & 56243.3 \\
n2      & 60001111003 & -  & -                  & 57.0  & 56243.8 \\
n3      & 60001111005 & -  & -                  & 61.5  & 56245.3 
\enddata
\end{deluxetable}

\suzaku XIS data are reduced using \texttt{aepipeline} in \textsc{heasoft}. We use the same observations as \cite{2017A&A...603A..50B} (summarized in Table \ref{tab:obs_log}). All the spectra are grouped to to have a minimum signal to noise ratio of 6 per bin, oversampling the detector resolution by a factor of 3.

In all the analysis, we use {\sc heasoft 6.25} and {\sc sas xmmsas\_20180620\_1732}. All spectral modeling is done in {\sc xspec 12.10.1} \citep{1996ASPC..101...17A}. For the spectral modeling we use $\chi^2$ statistics and calculate the errors on spectral parameters using the {\tt error} command. All the errors quoted in the text represent $1\sigma$ uncertainty.

For the timing analysis, we only use observations with exposures larger than 10 ks. These are marked with $(^t)$ in table \ref{tab:obs_log}. In order to obtain the longest and most continuous light curves possible, we select only good time intervals where the background rate between 10--12 keV is below 0.5 counts per second. This slightly relaxed from the 0.4 counts per second in the standard spectral extraction. We also use circular regions for the source extraction rather than the annular regions used when extracting the spectra. Pileup is unlikely to affect the timing data explored here. Light curves in the energies of interest are extracted by filtering on the pulse height PI values, then we use \texttt{epiclccorr} to apply both absolute (vignetting, bad pixels, chip gaps, Point Spread Function and quantum efficiency) and relative (deadtime, Good Time Intervals, exposure and background) corrections. All the light curves analyzed in this work are background-subtracted using the same extraction regions from the spectral analysis. 

\section{Spectral Analysis} \label{sec:spec}

\begin{figure}[t]
\epsscale{1.2}
\plotone{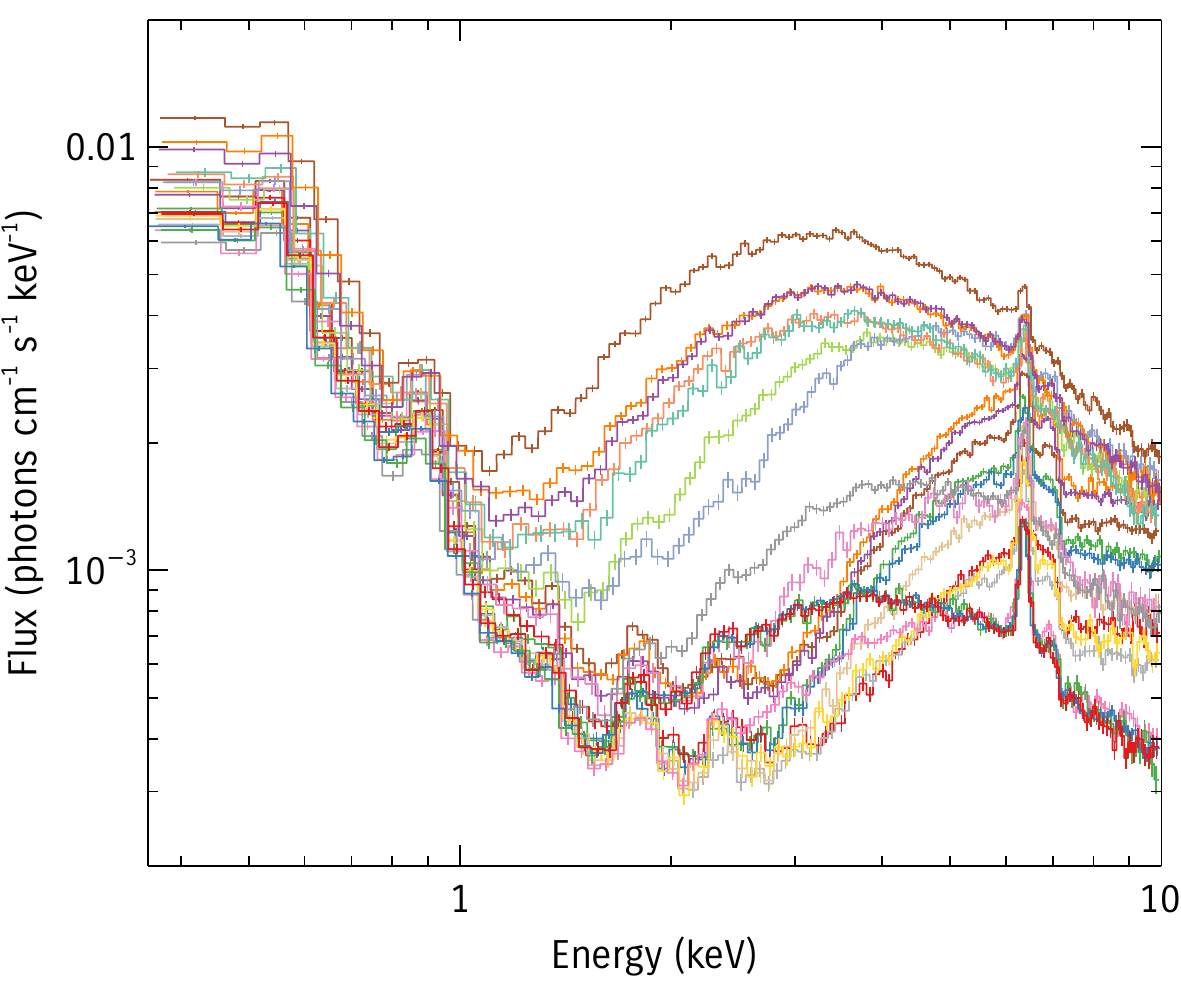}
\caption{The spectra from the EPIC-PN detector from the 22 observations described in Table \ref{tab:obs_log}. The spectra are divided by the effective area curve.\label{fig:show_spec}}
\end{figure}

\subsection{The General Shape}
\ngcfour is known for it complex spectrum. An illustration of the complexity of the available data is shown in Figure \ref{fig:show_spec}. The figure shows the spectra from all the 22 EPIC-PN camera exposures, where the data are divided by the response effective area at each energy. The plot shows that variations in the spectral shape can be due to a combination of intrinsic source variability (manifested in the changes between 8--10 keV) and to changes in the line of sight absorption (manifested as changes in the spectral curvature between 1--5 keV). We note also that the variability is most apparent above 1 keV. The band below 1 keV, which has been shown to contain a significant contribution from emission lines originating in a diffuse plasma, shows non negligible variability over the years. This suggests that photons from the central source leak through the absorber, a point we will discuss further in section \ref{sec:spec:toward_mod}.

\begin{figure}[t]
\plotone{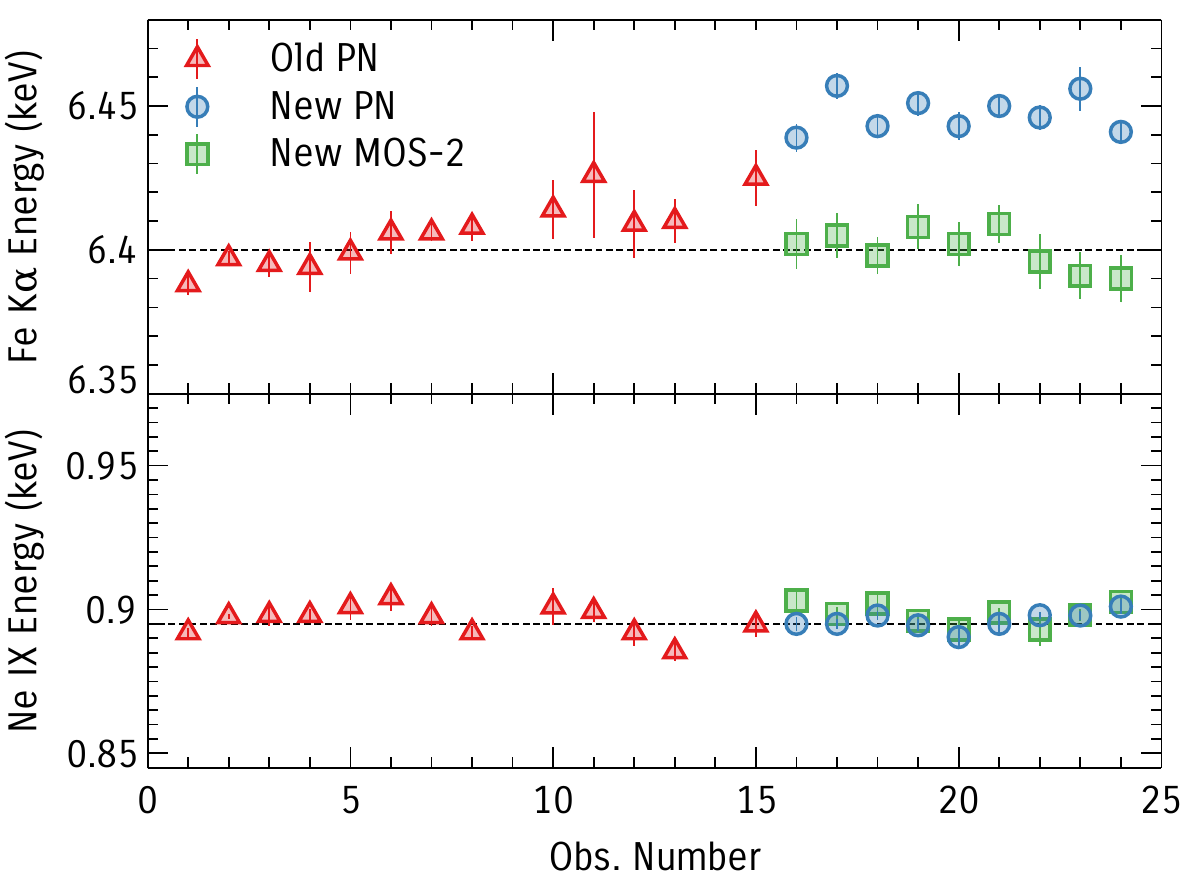}
\caption{Energy change showing the gain shift\label{fig:gline_shift} at the \FeK line energy. No shift is seen in the below 1 keV.}
\end{figure}

\subsection{EPIC-PN Gain Correction}
Before proceeding with the spectral analysis, and to check specifically for energy calibration of the EPIC PN, we measured the centroid energies of the strong \FeK line at 6.4 keV from all the 22 spectra. We fit the spectra between 6 and 7 keV with a power law model for the local continuum and a Gaussian line. The resulting energies of the line are shown in the top panel of Figure \ref{fig:gline_shift}. 

We show data from the old (observations 1--16) and new (observations 17--24) \xmm observations separately, and also from the EPIC-MOS-2. The plot shows that most of the points from the old observations are clustered around 6.4 keV, implying an origin in neutral \FeK emission. The new PN data all show a systematic shift of $>40$ eV, which is a factor of 4 higher than the standard energy scale calibration uncertainty of the PN (Guainazzi et al 2011,  XMM-SOC-CAL-TN-0083\footnote{http://xmm2.esac.esa.int/docs/documents/CAL-SRN-0322-1-0.ps.gz}). The fact that the simultaneous MOS data are also clustered around 6.4 keV, suggests that the discrepancy is due an incorrect gain calibration in PN. Gain is the conversion of the
charge signal deposited by a detected photon from detector units to energy in eV. An incorrect calibration causes photons to be assigned an energy that is different from their true energy. 

We also check for gain shifts in the band below 1 keV using the strong Ne {\sc ix} at 0.9 keV. The measured energies are shown in the bottom panel of Figure \ref{fig:gline_shift}. The figure now shows that the gain shift of $\sim 40$ eV seen at 6.4 keV is not present below 1 keV. 
In order to mitigate the gain shift problem at the \FeK energies, and to allow further spectral modeling, we use the \texttt{gain} model in \textsc{xspec} to apply a linear shift to the energies on which the response matrix and the area curve are defined. In order to obtain the values of the linear parameters, we exploit the rich emission line spectrum of \ngcfour below 2 keV observed simultaneously with the RGS and assume that the \FeK line is neutral. 

First, the RGS spectrum between 0.35--1.0 keV of every observation is modeled using a power law continuum and a sum of emission and absorption lines. Lines are added to the model until the fit null hypothesis probability is $p>0.05$. We then fix the parameters of the best fit model, and use them to model the PN spectra, allowing for a cross-normalization constant. This model anchors the gain model in the PN data in the 0.35--1.0 keV band. The anchor in the hard band is achieved by fitting the 6--7 keV band at the same time with a power law for the local continuum and a neutral reflection model ({\tt xillver}, setting the ionization parameter to 0). This models the line at 6.4 keV, and its Compton shoulder. We are here explicitly assuming the \FeK line is neutral. This is justified by the fact that there has been no evidence for it being ionized. Also, as we show in section \ref{sec:neutral_abs}, despite the Fe K edge energy showing indication of ionized absorption in some observations, it is not clear that the line is produced in the same absorbing material, and even if it is, the line energy still peaks at 6.4 keV at such low ionization. We note that the gain parameters are not sensitive to the model used to fit the line as we found by using a simple Gaussian line at 6.4 keV. More complex models such as {\it MYTORUS} \citep{2009MNRAS.397.1549M} also give similar results given the limited energy band used. We use {\it xillver} \citep{2013ApJ...768..146G} to be consistent with the spectral modeling in subsequent sections. 

The gain correction parameters are obtained by adding the linear {\tt gain} model and fitting for its slope and offset. The additional gain model improves the fit in all cases. The resulting slopes and offsets of the gain model for each observation are shown in Table \ref{tab:obs_log}. All the observations are tested in this manner. The gain correction is then applied to the response and area files which are used in subsequent analysis.

\subsection{Towards a Full Model}\label{sec:spec:toward_mod}
Before attempting to model all the complex spectral data of \ngcfour, we explore signatures of different components that contribute to shaping the spectrum in sections \ref{sec:soft_emission_lines}, \ref{sec:neutral_abs} and \ref{sec:ionized_abs}, then we present the full model in section \ref{sec:full_model}.

\begin{figure}[t]
\includegraphics[width=\columnwidth]{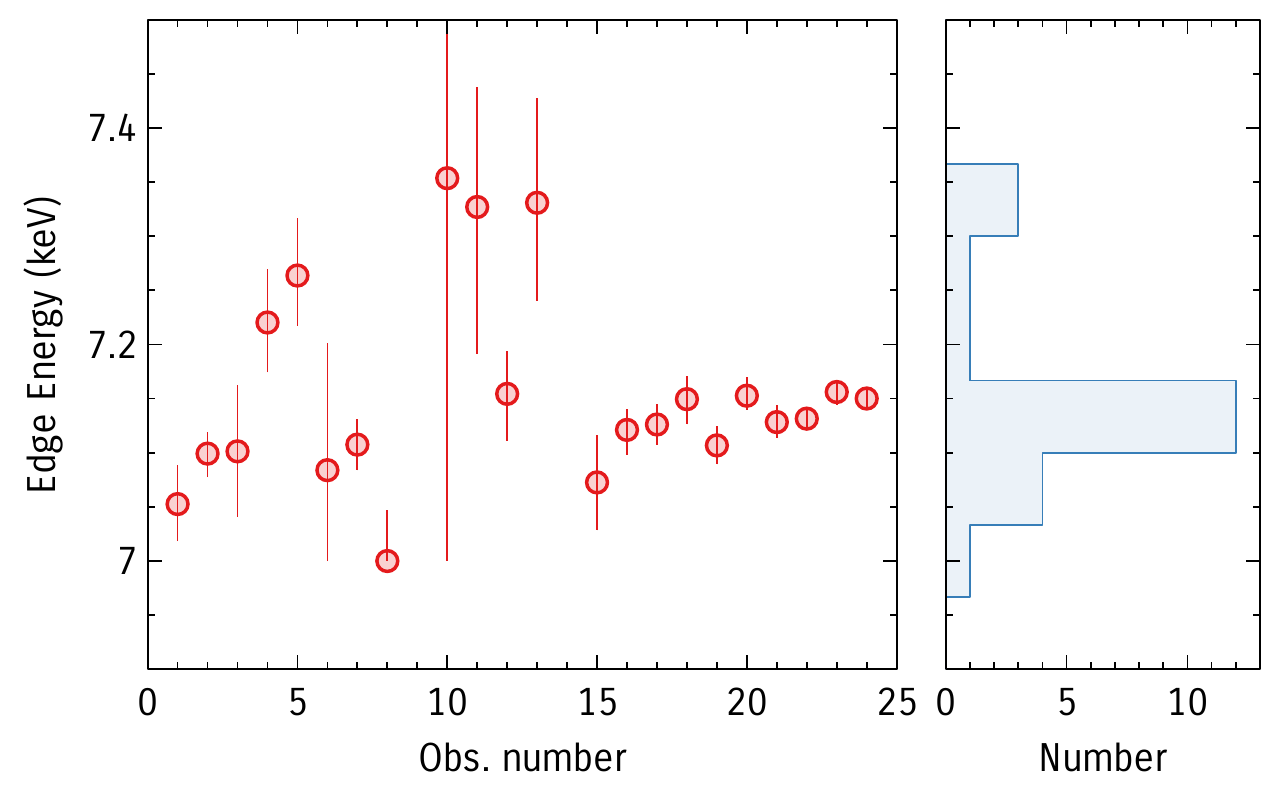}
\caption{Observed distribution of the Fe K edge energy. {\em Left:} A plot of the energy of the Fe K edge (in the source frame) for the different observations, modeled with a simple {\tt zedge} model. {\em Right:} A histogram of the observed edge energies.\label{fig:hard_edge}}
\end{figure}

\subsubsection{The Soft Emission Lines}\label{sec:soft_emission_lines}
The spectrum of \ngcfour below $\sim2$ keV is dominated by strong emission lines. \chandra images showed that a significant fraction of this soft emission originates in extended regions several hundred parsecs to 1 k-parsec in size, while the hard emission ($>2$ keV) is unresolved \citep{2001ApJ...563..124Y}. High spatial resolution emission line images of blended O {\sc vii}, O {\sc viii}, and Ne {\sc ix} show extended structures that are spatially correlated with the radio jet and optical O {\sc III} emission. The emission line spectrum contains a combination of photoionized and collisionally-ionized plasma \citep{2000ApJ...545L..81O,2011ApJ...736...62W}. For simplicity, we model this component with a Bremstruhlang model and a number of Gaussian functions.

The emission spectrum from the collisionally-ionized plasma is unlikely to vary - even on time scales of years - given its large size. The variability we see below $<2$ keV in Figure \ref{fig:show_spec} suggests that a fraction of the soft emission has to originate in a small region. As we will discuss in section \ref{sec:full_model}, the soft variability can be attributed to some nuclear emission that leaks through the absorber, or to variability in the emission spectrum originating in the photoionized plasma.

\subsubsection{Neutral Absorption}\label{sec:neutral_abs}
The variability in the shape of the spectra in Figure \ref{fig:show_spec} around 3 keV, and the depth of the strong absorption edge at $\sim7$ keV both point to absorption as the main contributor to shaping the spectrum.
Neutral absorption produces a sharp edge at 7.1 keV, and as the ionization increases, the energy shifts to $\sim7.5$ keV \citep{2002ApJ...577L.119P}. Figure \ref{fig:hard_edge} shows the distribution of measured edge energies from individual observations. These are found by fitting the spectra between 6--10 keV with a power law, a redshifted edge ({\tt zedge}) and two Gaussian functions at 6.4 and 7.06 for the \FeK and Fe K$\beta$ lines. The distribution shows that for most observations, the edge is consistent with neutral absorption. In a handful of observations, the edge energy is at $\sim 7.3$, suggesting absorption by gas with an ionization parameter of $log(\xi)\sim1$ ($\xi$ is the ratio of the ionizing X-ray flux to gas density $\xi=4\pi F_x/n$). We note that the highest edge energies corresponds to the observations with the highest unabsorbed flux, suggesting that the absorber is being ionized by the nuclear emission. We discuss this point further in section \ref{sec:discuss:wa}.

\begin{figure*}
\includegraphics[width=\textwidth]{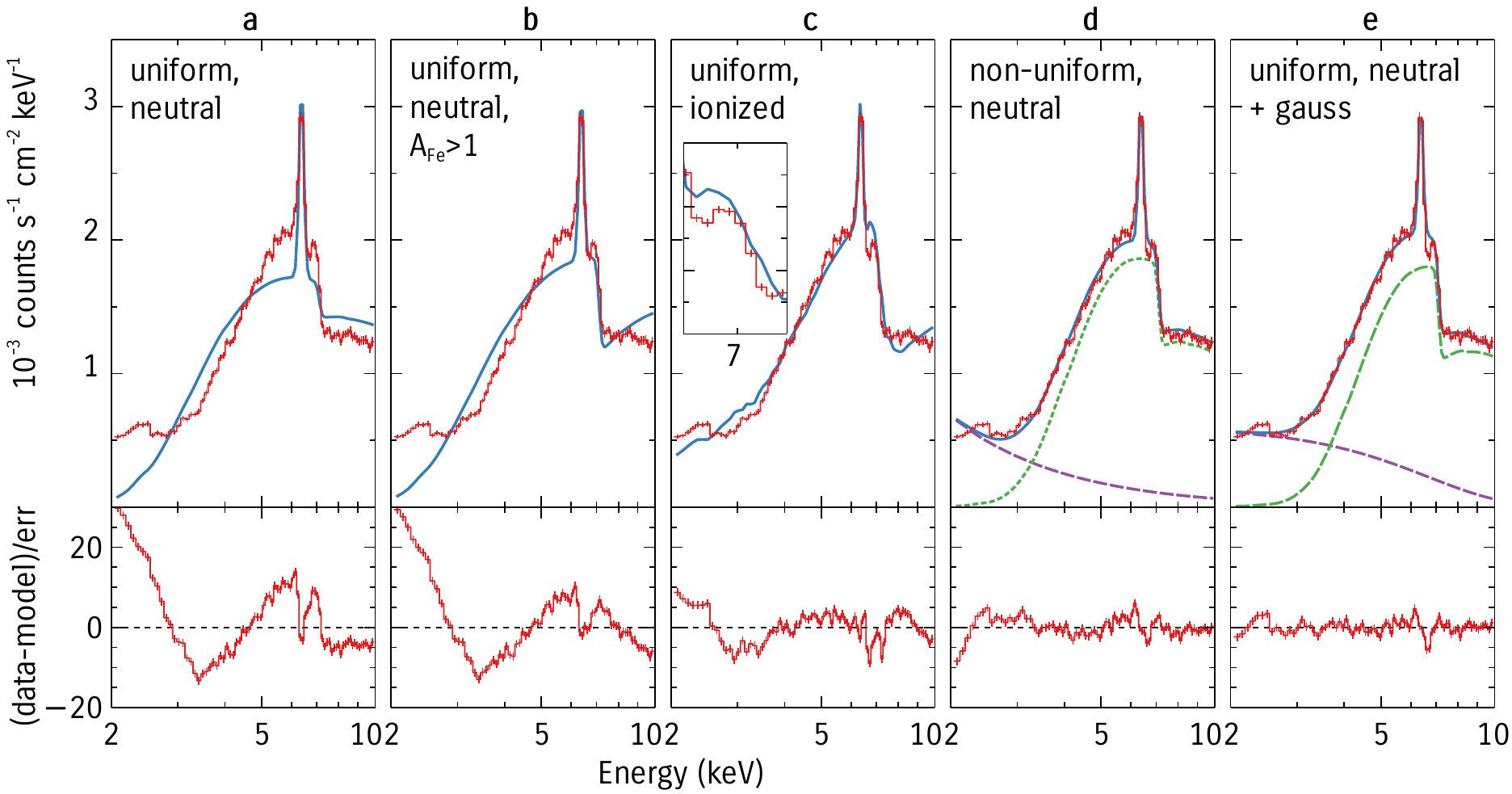}
\caption{Spectral modeling of the hard spectrum of \ngcfour. The plots show the spectrum from observation 24 where the strong neutral absorption is prominent. The top panel shows the spectral data and the best fit model. The residuals in each case are shown in the bottom row. The models are as follows. {\bf a:} a power law seen through a uniform neutral absorber ({\tt TBabs}). {\bf b:} Similar to {\bf a}, but allowing for super-solar iron abundances. {\bf c:} Using an ionized instead of a neutral absorber ({\tt zxipcf}). The inset zooms onto the iron edge. {\bf d:} Similar to {\bf a}, but the absorber is patchy, allowing some of the power to leak through to model the excess at $\sim 3$ keV. {\bf e:} Similar to {\bf d}, but using an extra components (a guassian function) to model the excess at 3 keV.\label{fig:hard_pc}}
\end{figure*}

We illustrate some of the properties of the neutral absorber further by focusing on one spectrum where the absorption is prominent, observation 24. This spectrum shows a strong and sharp Fe edge, and strong curvature between 3--6 keV. Figure \ref{fig:hard_pc} shows different models that attempt to reproduce both of these features. Panel {\bf a} shows the case of a uniform neutral absorber (\texttt{ztbabs}). The depth of the edge requires a large column ($N_{\rm H}\sim3\times10^{23}$ cm$^{-2}$), but such a column produces a sharp drop below 5 keV, that is well below the data. A possible remedy that has been suggested before is that the absorber has super-solar abundance \citep[e.g.][]{2003MNRAS.345..423S}. This produces a stronger edge for the same lower energy curvature. Panel {\bf b} shows that the data remain well above the model prediction between 2--3 keV. This super-solar abundance would also predict a much stronger line, which is not shown here. Using an ionized absorber (panel {\bf c}) allows some of the continuum emission to leak through, producing some of the 2--5 keV curvature, but fails to model the edge correctly (inset in panel {\bf c}). This suggests that either the absorber is not uniform or that there is a separate component that has significant contribution around 3 keV.

The first option is illustrated in panel {\bf d}, where we assume that some of the continuum emission leaks through the absorber (partial covering absorber). This model accounts for both the edge and curvature, though some residuals remain. In the second option, we explored adding another power law component or a broad Gaussian function. Adding a power law is mathematically similar to the non-uniform absorber case, and such a model can be due to scattering, which is commonly observed in Seyfert-2 galaxies \citep{1997ApJ...488..164T,2019arXiv190411028K}. For the Gaussian function, we found that it peaks at 2--3 keV. This is unlikely to be a redshifted iron line, as the required high redshift imply that the Compton reflection hump would also be shifted to the XMM-PN band, which is not observed. When the Gaussian component is fitted to all the spectra, we find that its flux is tightly correlated with the hard component. {\em We therefore conclude that although a separate component cannot be ruled out, the simpler hypothesis is that the additional component at 2--3 keV is the same hard component leaking through the absorber or being scattered by surrounding distant gas}. In the rest of this work, we will refer to this model as the secondary power law component, but it can physically be produced by scattering, non-uniform absorption, or by an additional component that dominates at 3 keV.

\subsubsection{Ionized Absorption}\label{sec:ionized_abs}
The strong emission lines in the soft band mask bound-free absorption from O {\sc vii}, O {\sc vi}, Ne {\sc ix} and Ne {\sc x}, which are the typical signatures of warm absorbers \citep{1984ApJ...281...90H,1997MNRAS.286..513R,2005A&A...431..111B}. Signatures of ionized absorption in this source, therefore, have come from studying the weaker and less abundant elements such as Mg, Si and S \citep{2005ApJ...633..693K,2016ApJ...833..191C}.

\begin{figure}
\includegraphics[width=\columnwidth]{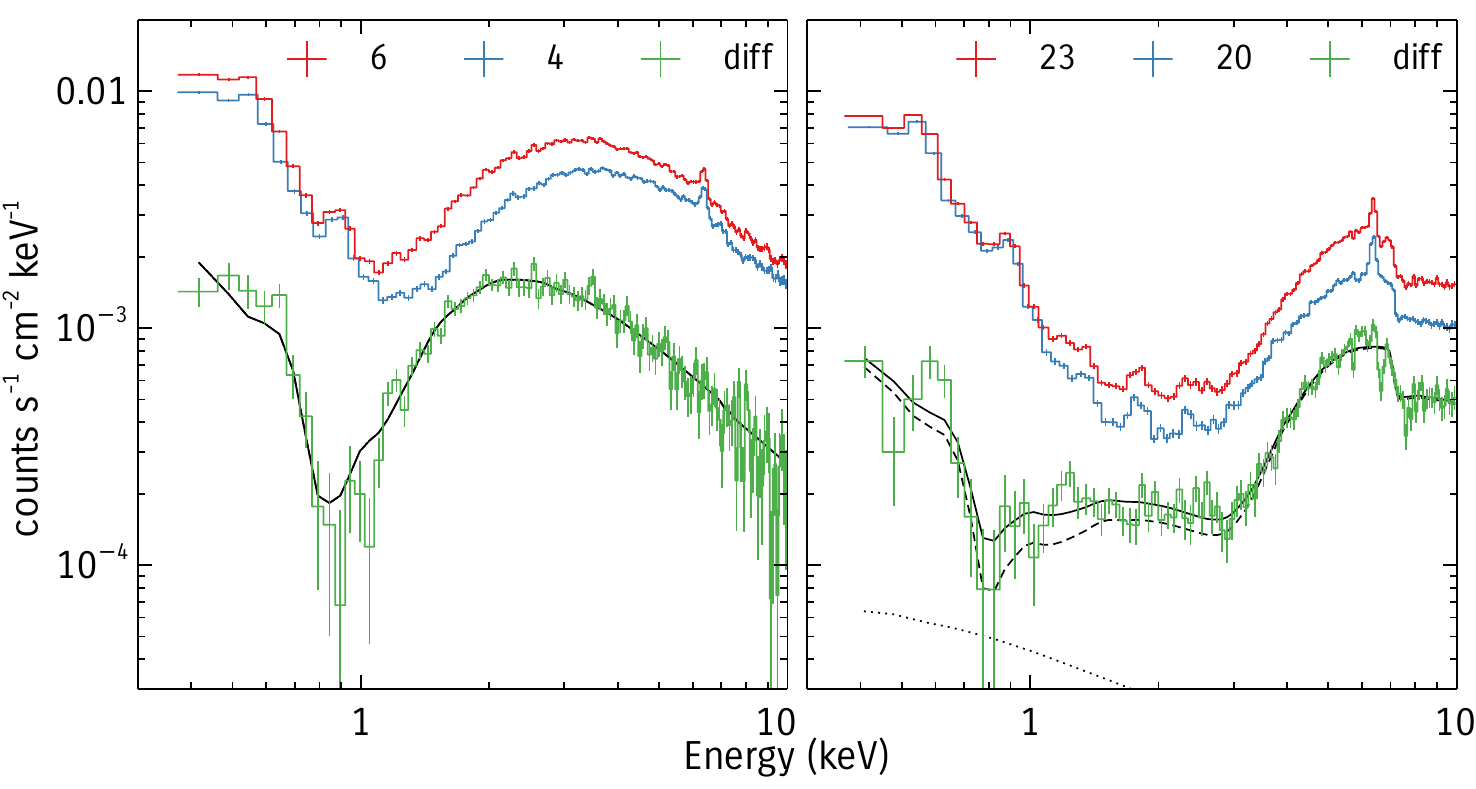}
\caption{High (red) and low (blue) flux spectra and their difference (green) for two pairs of observations. The left panel shows the spectra when the absorption is low ($N_{\rm H}\sim6\times10^{22}$; observations 6 and 4) and the right panel shows the spectra when the absorption is high ($N_{\rm H}\sim3\times10^{23}$; observations 23 and 20). The difference spectra show the shape of the variable component, which has strong troughs at $\sim0.8$ keV, indicative of ionized Oxygen.\label{fig:diff_wa}}
\end{figure}

Evidence for strong, ionized absorption would be more apparent at lower energies. Although the 0.3--1 keV band is dominated by emission lines, we can reduce their effect by using difference spectra, where a low flux spectrum is subtracted from a high flux spectrum. Any constant emission components (here in the 0.3--1 keV band) are subtracted out, and only the variable spectrum remains. This is a simple and model-independent way to find the variable spectrum. We plot in Figure \ref{fig:diff_wa} the difference spectra for two pairs of observations. For each pair, high and low flux spectra are chosen to have comparable neutral absorption columns, as inferred by the Fe K edge and the 3--5 keV curvature, so that the non-linear effect of absorption on the difference spectra are minimized. The high and low flux observations are also chosen to be close in time (a few days at most) to avoid any gain shift artifact when the same response file is used in constructing the difference spectrum. The spectra are produced by first loading the high flux spectrum into {\sc xspec} and then loading the low flux spectrum as background. The two pairs are representative of spectra where the source is seen through low (left; $N_{\rm H}\sim6\times10^{22}$; observations 6 and 4)) and high (right; $N_{\rm H}\sim3\times10^{23}$; observations 23 and 20) neutral absorption columns.

The left panel of Figure \ref{fig:diff_wa} shows that by subtracting the constant (mostly diffuse) emission lines in the case of the spectra 4 and 6, we can see that the variable component resembles a classical warm absorber spectrum, with a trough at $0.7-0.8$ keV, consistent with the bound-free absorption from Oxygen. The trough at 0.7 keV can be modeled with an absorber with an ionization parameter of $log \xi\sim1$, as inferred from the Fe K edge energy, and an equivalent Hydrogen column density of $N_{\rm H}\sim3\times10^{22}$ cm$^{-2}$. A neutral absorber alone cannot reproduce the trough.

The 22--20 difference spectrum in the right panel of Figure \ref{fig:diff_wa} is more complex.  The flat part at 2--3 keV is likely a result of the secondary power law component discussed in Section \ref{sec:neutral_abs}. Also, in addition to a simple ionized absorber similar to the 6--4 case, the spectrum below 1 keV requires an additional power law to capture the general spectral shape. Some of the soft emission lines show up in the difference spectrum too, suggesting that part of the emission spectrum is variable. The diffuse collisionally ionized plasma is unlikely to be variable given its large size, and the variability we observed is most likely due variability of the photoionzed gas that is located at sub-parsec scales. It is then likely that the collisionally-ionized plasma is constant and produced on kpc scales, which appears to be partially  correlated with the radio jet and the optical O III emission, while the variable photonionized part originates at sub-parsec scales. Separating the two components is not possible given the PN energy resolution, and may be requires detailed modeling of higher resolution RGS data, which is beyond the scope of this work.

The 22--20 difference spectrum also shows that the \FeK line is variable. A strong absorption line at 6.7 keV is also visible. This is likely due to Fe {\sc xxv} produced in highly ionized gas ($log\xi>3$). This absorption line - along with the trough at 0.7 keV - indicate that there are at least {\em two} ionized absorbers in the line of sight: low ($log\xi\sim1$) and high ($log\xi>3$) ionization components.

\subsection{Modeling All the Spectra}\label{sec:full_model}
In this section, we model all the 22 spectra in the 0.3--10 keV band using a model that takes into account the soft emission, the neutral and ionized absorption discussed in section \ref{sec:spec:toward_mod}.

\begin{figure}[t]
\centering
\includegraphics[width=0.85\columnwidth]{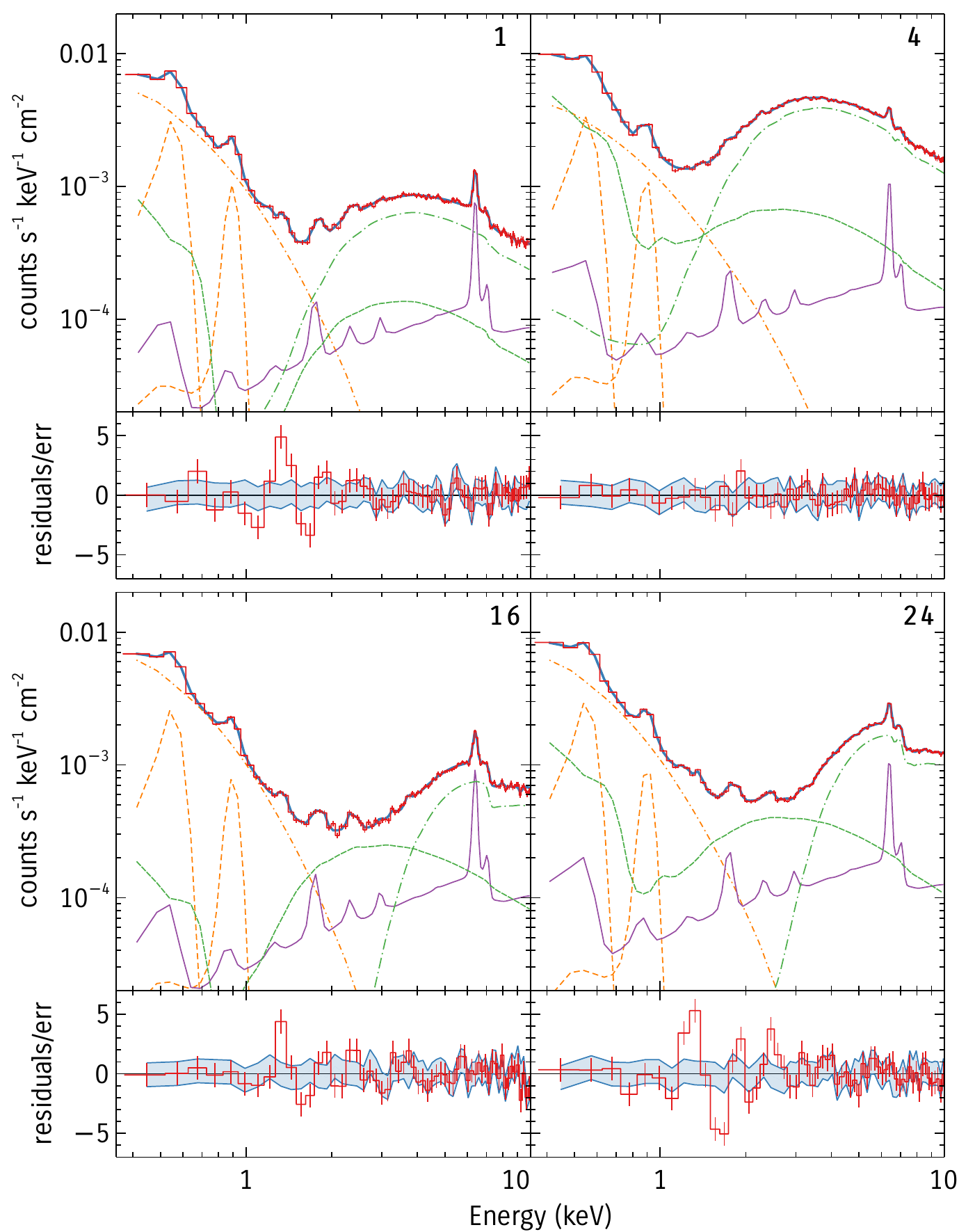}
\caption{Results of model fitting shown for a representative subset of the spectra using observations 1, 4, 16 and 24, that span the observed fluxes and absorption column densities. {\em Spectral row}: The data is shown in red, the total model in blue. The primary (absorbed) power law is shown in dot-dashed green and the secondary is shown in dashed green. The distant reflection component is in purple, while the soft components are shown in orange. {\em Residuals row}: The red points are the residuals for the model that include only the strongest soft lines. The blue bands are the residuals when additional emission lines are added to the soft band (See text for details). A similar plot showing all the spectra is shown in Figure \ref{fig:apdx:fit4_all}. \label{fig:fit4_4spec}}
\end{figure}

\begin{figure*}[t]
\centering
\includegraphics[width=\textwidth]{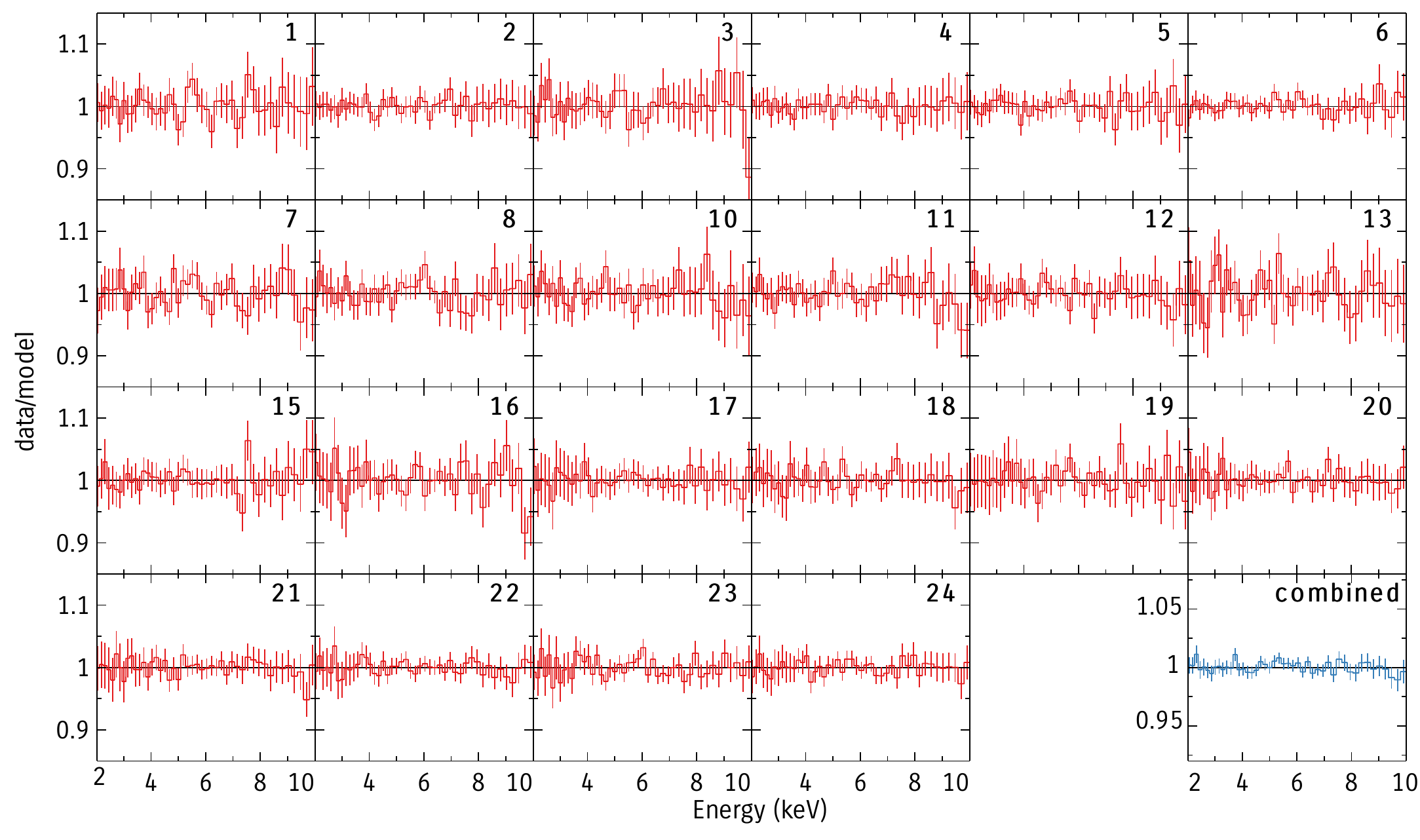}
\caption{Data to model ratio for the final best fit model. The combined residuals are produced by averaging the counts and the best fitting model and taking their ratio. Note that the combined residuals plot has a different y-axis scale.\label{fig:fit4_4c_rat}}
\end{figure*}

The primary continuum is modeled with a power law ({\tt powerlaw} in {\sc xspec}; \xsm{po}{h} henceforth). 
The warm absorption is modeled with \xsm{zxipcf}{} \citep{2008MNRAS.385L.108R}, and we refer to them as \xsm{zxipcf}{l} and \xsm{zxipcf}{h} for the low and high ionization components respectively. 
The secondary component that dominates at 2--3 keV is assumed to be a power law with a photon index that equals the index of the primary power law, and it is modeled as partial coverer along with the neutral absorption using \xsm{TBpcf}{} \citep{2000ApJ...542..914W}. As we discussed in section \ref{sec:spec:toward_mod}, this can be a physical partial covering component, but it does not have to be.
We also include uniform neutral absorption from our galaxy using \xsm{TBabs}{}.
The soft spectra below 1 keV are modeled with with a Bremsstrahlung component (\xsm{brems}{s}) and two Gaussian emission lines at $\sim 0.57$ and $\sim 0.9$ keV to model the strong O \textsc{vii} (\xsm{zgauss}{O}) and Ne \textsc{ix} (\xsm{zgauss}{Ne}) respectively. Using a power law instead of \xsm{brems}{s} gives similar results. The strong \FeK line is modeled with the reflection model \xsm{xillver}{}, which also models the scattering shoulder of the line at $\sim6.3$ keV and the Fe K$\beta$ line. The photon index of \xsm{xillver}{} is linked to that of \xsm{po}{h} and we assume solar abundances and an inclination of 30$^{\circ}$. The final model has the following {\sc xspec} form:

\begin{tt}
\small\noindent
TBabs(\xsm{zxipcf}{l}(\xsm{zxipcf}{h} * \xsm{TBpcf}{} * \xsm{po}{h}) \\
+ xillver + \xsm{brems}{s} + \xsm{zgauss}{Ne} + \xsm{zgauss}{O})
\end{tt}

The order of the absorption components is justified by the difference spectra in Figure \ref{fig:diff_wa}. The fact that we see a trough at 0.7 keV in the variable spectrum indicates that the low ionization warm absorber has to be outside the neutral absorber producing the strong Fe K edge, otherwise the neutral absorption would block any emission $<1$ keV.

The same model was fitted to all the spectra. The best fit model and fit residuals for a subset of the spectra are shown in Figure \ref{fig:fit4_4spec} (A similar plot showing all the spectra is shown in Figure \ref{fig:apdx:fit4_all} in the appendix). The spectra selected for Figure \ref{fig:fit4_4spec} are representative and span the observed flux levels and absorption column densities. The residuals to the fit are shown in red in the lower panel of each plot. The fits are not statistically acceptable for any spectrum mostly because of the emission lines below $<2$ keV that are not modeled yet.

Next, we add Gaussian functions to model these emission lines in the soft band. Gaussian lines with energies between 0.3--4 keV are added to the model one at a time, until the added line does not provide any significant (at the 90\% level) improvement to the fit. The overall fit improves significantly for all the spectra. The residuals for this fit are shown in Figure \ref{fig:apdx:fit4_4a}. Absorption lines can seen in the 2--5 keV range that are not modeled by the two warm absorption layers we included in the model. This includes lines at 2.3 keV (Si {\sc xiv}), 4.9 keV (Ca {\sc xx}), and 2.9 keV. The lines are weak in individual observations, but the fact some are seen in multiple observations suggests they are real. These absorption features are modeled with Gaussian lines. 

Some observations (e.g. 10, 11, 18) also show an absorption line at 9.2 keV. The feature is very apparent in the combined residuals plot in the same figure. If this is due to Fe {\sc xxvi}, then an outflow velocity of $0.5c$ is implied. The significance of this feature in individual observations is low, but the fact that it is present in at least 5 observations may also suggest that it is real. The feature is very close to the Fe {\sc xxvi} K edge at 9.28 keV at zero velocity, so it could also be a weak part of the warm absorption system. If adding an absorption line at 9.2 keV provides a significant (at least 90\% confidence) to the fit, we include it in the model. The same is done also for an apparent emission line at 7.5 keV (apparent for example in observations 6 and 8 in Figure \ref{fig:apdx:fit4_4a}), likely due to Ni K$\alpha$\footnote{Note that because these lines are not the primary focus of this work, we do not make strong statements about the significance. The 90\% confidence level we use does not account for the number of trials, but we include them because they appear to have an affect on the combined residuals above 7.5 keV.}.

The residuals of the best fit model after accounting for these weak features is shown in blue in the bottom panel of each spectrum in Figure \ref{fig:fit4_4spec} (and Figure \ref{fig:apdx:fit4_all} for all the spectra). A summary of the main fit parameters is shown in Table \ref{tab:fit_indiv}. The model provides an very good fit to all the spectra as indicated by the fit statistic and null hypothesis probability in the last two columns of Table \ref{tab:fit_indiv}.

A closer examination of the residuals in the 2--10 keV is shown in Figure \ref{fig:fit4_4c_rat}. The plot do not suggest any remaining strong features in the 2--10 keV, including a relativistic broad iron line. The combined residuals rule out any additional features with contribution larger than $\sim 1\%$. The only observation where the model is rejected is observations 23 ($p=10^{-4}$; see Table \ref{tab:fit_indiv}), this is mostly due to two apparent absorption lines at 5.6 and 7.4 keV likely due to a more complex absorption system that is not captured by our two {\tt zxipcf} components. Adding a broad emission Gaussian function at $\sim 6$ keV improves the fit only $\Delta\chi^2=6$ for 3 degrees of freedom, which is not significant. For a handful of observations, the null hypothesis probability suggests over-fitting, and these spectra may be fitted with simpler models. Our strategy, however, was to include all components required by all the spectra. This allows us to obtain limits on components that are not strictly required in individual spectra, but that are present in other epochs. In the following section, we compare these conclusions to previous studies which implied the presence of a relativistic component.

We note that our best model has some degeneracy between some of the absorption parameters. It is conceivable that other variants of the model are possible, such as a partial covering warm absorber, partial plus full covering neutral absorbers, plus a weaker warm absorber. All these models are effectively similar in general, and distinguishing between them may require including the higher resolution RGS data. Even then, the fact that most of the strong warm absroption signatures are masked by the emission spectrum below 2 keV may make that a challenge.


\subsection{Relativistic Reflection}\label{sec:relref}
In sections \ref{sec:neutral_abs} and \ref{sec:ionized_abs}, we showed that at least two layers of warm absorption and one non-uniform neutral absorber are required by the data. In section \ref{sec:full_model}, we showed that all the spectra can be described by a model that accounts for these effects, and that additional components, specifically a relativistic reflection component, are not required by the data. 

\begin{figure*}
\centering
\includegraphics[width=0.8\textwidth]{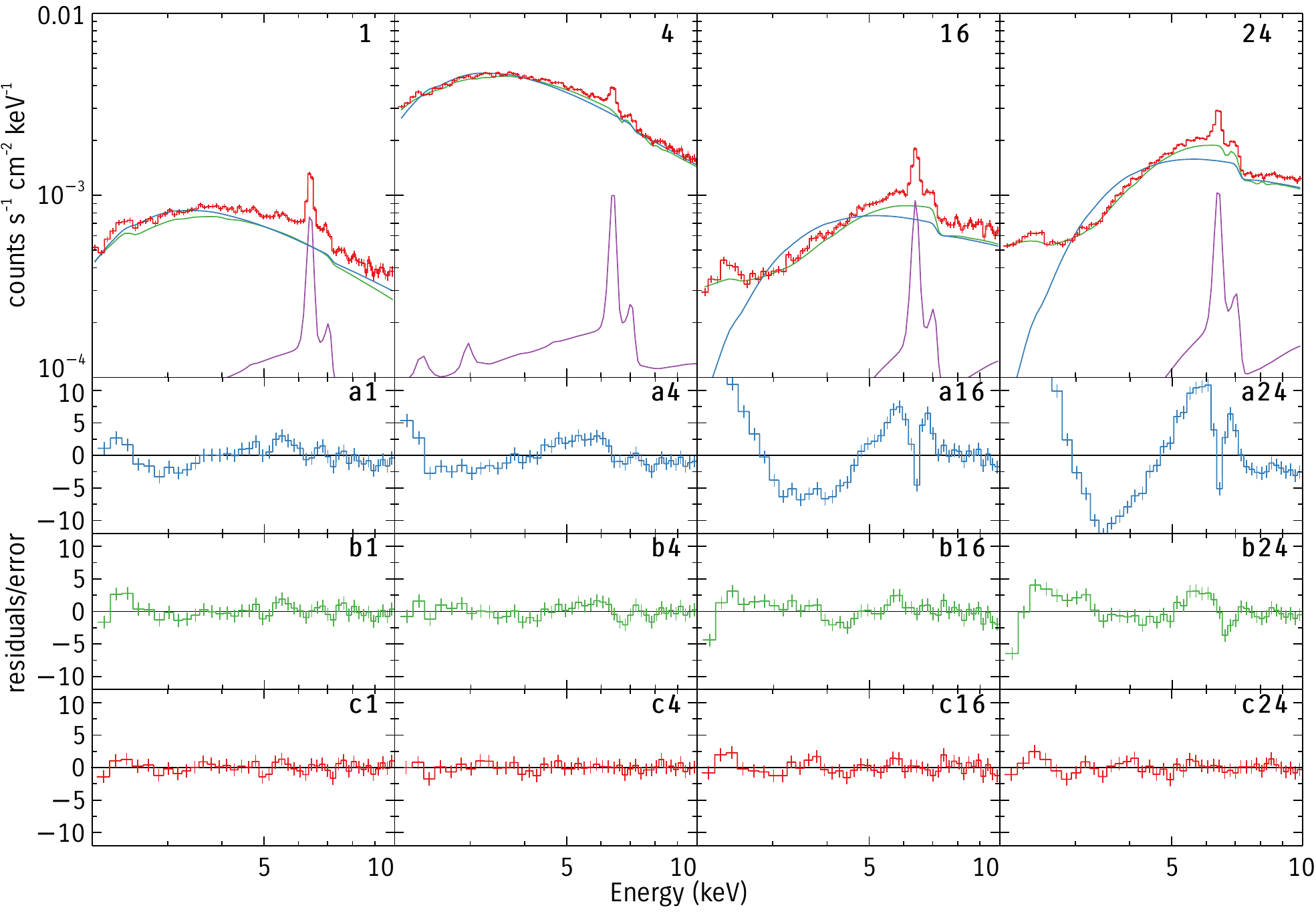}
\caption{Applying the model from \citetalias{2012MNRAS.422..129Z} (observations 1 and 4 in the 1st and 2nd columns) to the new data (observations 16 and 24 in the 3rd and 4th columns). The first row shows the spectral data and models. The purple line is the {\tt xillver} model. The blue line is power law model with neutral uniform absorption, and its residuals are shown in the 2nd row (labeled a1--a24). The green the line is the same model but making the absorber partial (i.e. allowing some emission to leak through the absorber), and the residuals are shown in green in the 3rd row (labeled b1--b24). The last row shows the residuals after adding a warm absorber.
\label{fig:z12}}
\end{figure*}

The presence of relativistic reflection in the spectra of \ngcfour have been reported in several studies (\citetalias{2012MNRAS.422..129Z}; \citealt{2015ApJ...806..149K, 2017A&A...603A..50B}). In \citetalias{2012MNRAS.422..129Z}, we focused on modeling the 2--10 keV spectra for subset of the \xmm data presented here. \cite{2015ApJ...806..149K} modeled a simultaneous \suzaku and \nustar dataset taken in 2012, focusing on the 2.5--80 keV. \cite{2017A&A...603A..50B} modeled all the \suzaku, \nustar and \xmm data observed prior to 2015. Their \xmm data corresponds to observations 1--15 in Table \ref{tab:obs_log}.

We revisit the modeling in these works in light of the most recent observations. In \citetalias{2012MNRAS.422..129Z}, we found that a model consisting of an absorbed power law and distant reflection leaves broad residuals that peak around the \FeK line. These residuals are shown in panels a1 and a4 in Figure \ref{fig:z12} for observations 1 and 4 in Table \ref{tab:obs_log}, which were used as illustrative examples in \citetalias{2012MNRAS.422..129Z}. The residuals in this case do resemble a broad Fe line, leading to the modeling in \citetalias{2012MNRAS.422..129Z}. Figure \ref{fig:z12} also shows in panels a16 and a24 similar plots from the latest observations 16 and 24 respectively. These observations are characterized by a higher absorption column density. The curvature in the residuals is stronger around the iron line, and also the spectrum rises steeply below 4 keV. The relativistic reflection model, which fit the line-like features in the old data (observations 1 and 4) cannot model the stronger feature in the new data (observations 16 and 24).

As we discussed in section \ref{sec:neutral_abs}, the strong curvature is likely due to an additional component, either due to nonuniform absorption in which a fraction of the primary source leaks through the absorber, or due to scattering of the nuclear emission by the circumnuclear gas. We illustrate this further by comparing the residuals from two models. The 3rd row of Figure \ref{fig:z12} (b1--b24) shows the residuals when a separate power law model is added to the spectra to account for the excess below 4 keV, assuming it has a photon index that equals that of the primary power law. This component accounts for some of the residuals resembling a broad line, with some residuals left at 2--3 and 6--7 keV, which are strongest again in the new data. These are reproduced by including the two ionized absorption components discussed in section \ref{sec:ionized_abs}. The 4th row of Figure \ref{fig:z12} (c1--c24) shows the residuals after adding these two components. The model now is the same as that already discussed in section \ref{sec:full_model}.

\begin{figure}[h]
\centering
\includegraphics[width=\columnwidth]{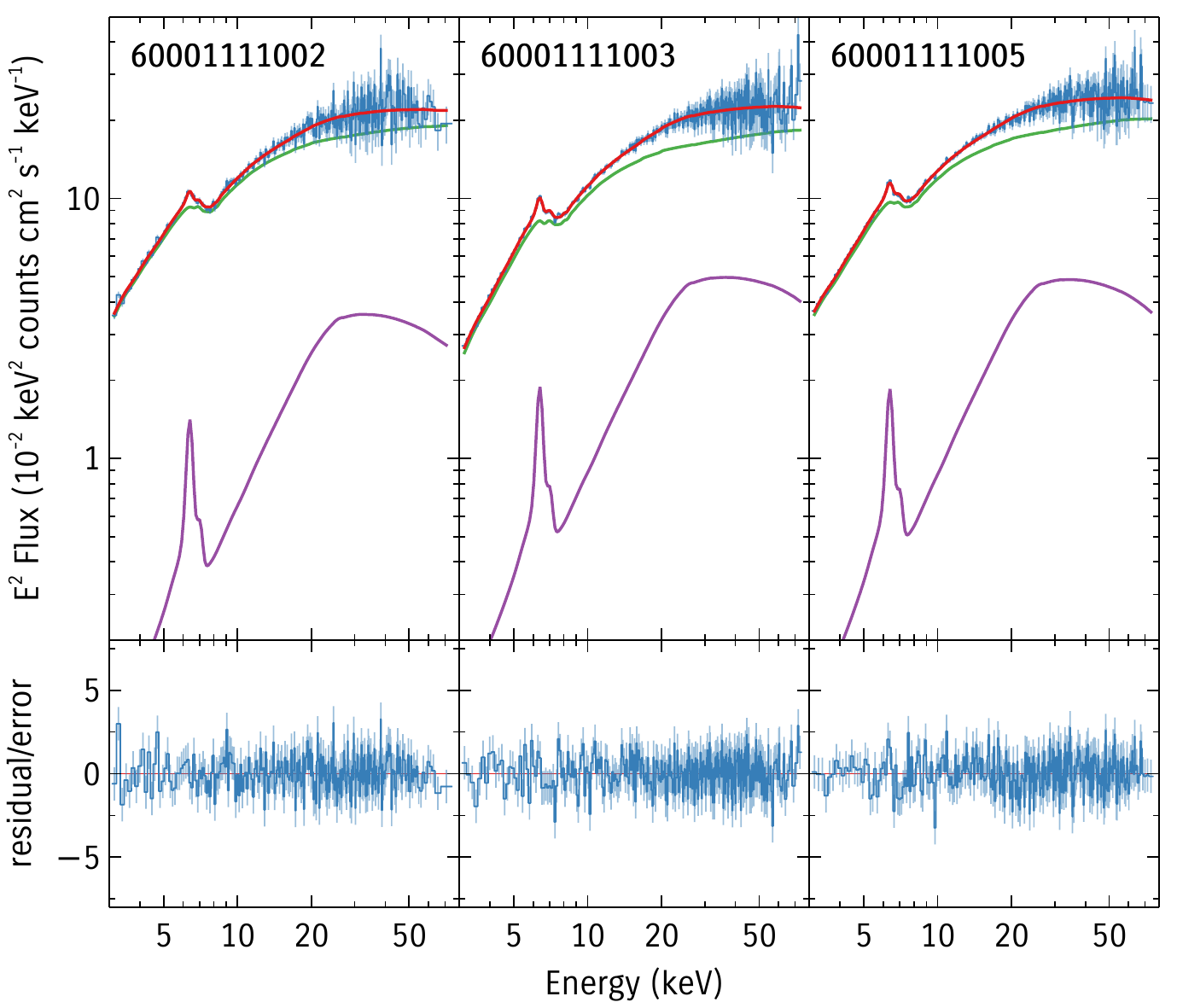}
\caption{The result of applying the model discussed in section \ref{sec:full_model} to the \nustar data. The observation ids are shown in the figure. The absorbed power law  (\xsm{po}{h}) and \xsm{xillver}{} are plotted along with spectral data and total model in the top panels, while the residuals are shown in the bottom panel.
\label{fig:nustar}}
\end{figure}

\cite{2017A&A...603A..50B} found that after including a partial covering model to account for the 2--3 keV spectral flattening, broad residuals remain\footnote{The model used in \cite{2015ApJ...806..149K} is similar to \cite{2017A&A...603A..50B} so we focus on the latter.}, and they are modeled with two lamp-post reflection models\footnote{Note that there is some inconsistency with modeling in regards to the Fe abundance. The abundance in the reflection model is a free parameter, with the best fit being 2.5 solar, while the absorption components have solar abundance. A non-solar abundance in the absorption models affects significantly the Fe K edge and the spectral curvature at 4 keV.}. Their strongly blurred reflection model (LP$_1$) is {\em only} required in one \suzaku observation out of 3 \suzaku and 7 \xmm observations. Those \suzaku data showed a significant discrepancy with the simultaneous \nustar observations at $<5$ keV where the LP$_1$ model has significant contribution. It is by discarding the \nustar data $<5$ keV and modeling only the \suzaku spectra that the LP$_1$ model is constrained.

\cite{2017A&A...603A..50B} also found that a second relativistic component (LP$_2$) was required by the data. This model produces only a weakly-broadened iron line, and it is not clear that such broadening is due to strong relativistic effects. The narrow \FeK line in \cite{2017A&A...603A..50B} was modeled with a Gaussian line, not accounting for the photons scattered before escaping the emission region, which produces the Compton shoulder at $\sim 6.2$ keV. Additionally, modeling of the higher resolution \chandra data \citep{2018arXiv180807435M}, suggests that the narrow \FeK line is asymmetric. Not accounting for these effects leaves a flux excess around the narrow \FeK line in the \suzaku data that appears like LP$_2$ in \cite{2017A&A...603A..50B}. We show in Figure \ref{fig:nustar} the result of fitting the model discussed in section \ref{sec:full_model} to only the \nustar data. No strong residuals remain after fitting an absorbed power law and a distant reflector. The excess in the data at 20--30 keV attributed previously to relativistic reflection is accounted for by the distant reflector.


\subsection{The Distant Reflector}\label{sec:narrow_lag}
Using the fits presented in section \ref{sec:full_model} and summarized in Table \ref{tab:fit_indiv}, we focus in this section on the distant reflector, characterized primarily by the strong narrow \FeK line at 6.4 keV. We find that the narrow \FeK line is {\em variable} between observations. Figure \ref{fig:cont_line}-a shows the variations of the line flux compared to the primary continuum. The continuum flux is measured between 7--10 keV, which is the observed part of the ionizing continuum. The line flux is measured between 6.1--6.7  keV. Note that what we call line flux also includes emission from the scattered Compton shoulder and a small contribution from the reflection continuum in the 6.1--6.7 keV band. Modeling the line with a single Gaussian function gives similar results but the line fluxes are smaller by $\sim 0.2$.

First, Figure \ref{fig:cont_line} shows that the \FeK line flux is variable on the time-scales probed by the data (days to years), and that it is highly correlated with the unabsorbed intrinsic flux (Spearman rank correlation coefficient $r=0.80$, $p=4\times10^{-6}$). A strong correlation implies that the line is \emph{responding} to the continuum variability. The line flux appears to be uncorrelated with the absorption column density $N_{\rm H}$ ($r=-0.14$, $p=0.73$; Not shown).

The fractional change (max-min)/(max+min) in the continuum and line fluxes are 0.7 and 0.3 respectively, suggesting that about half of the line flux is responding. This is a lower limit as it assumes the line at the lowest flux spectra contains only the constant component, which might not be the case, suggesting that a significant fraction of the line is variable.

\subsubsection{Constraint on the Narrow \FeK Delay from the Scatter}\label{sec:lag_scatter}
The existence of a correlation is a clear indication that the line responds to changes in the continuum. We attempt here to constrain the time delay between the line and the continuum using the scatter in the correlation. In section \ref{sec:lag_javelin}, we will directly measure the lag given some assumptions about the continuum variability.

\begin{figure}
\includegraphics[width=\columnwidth]{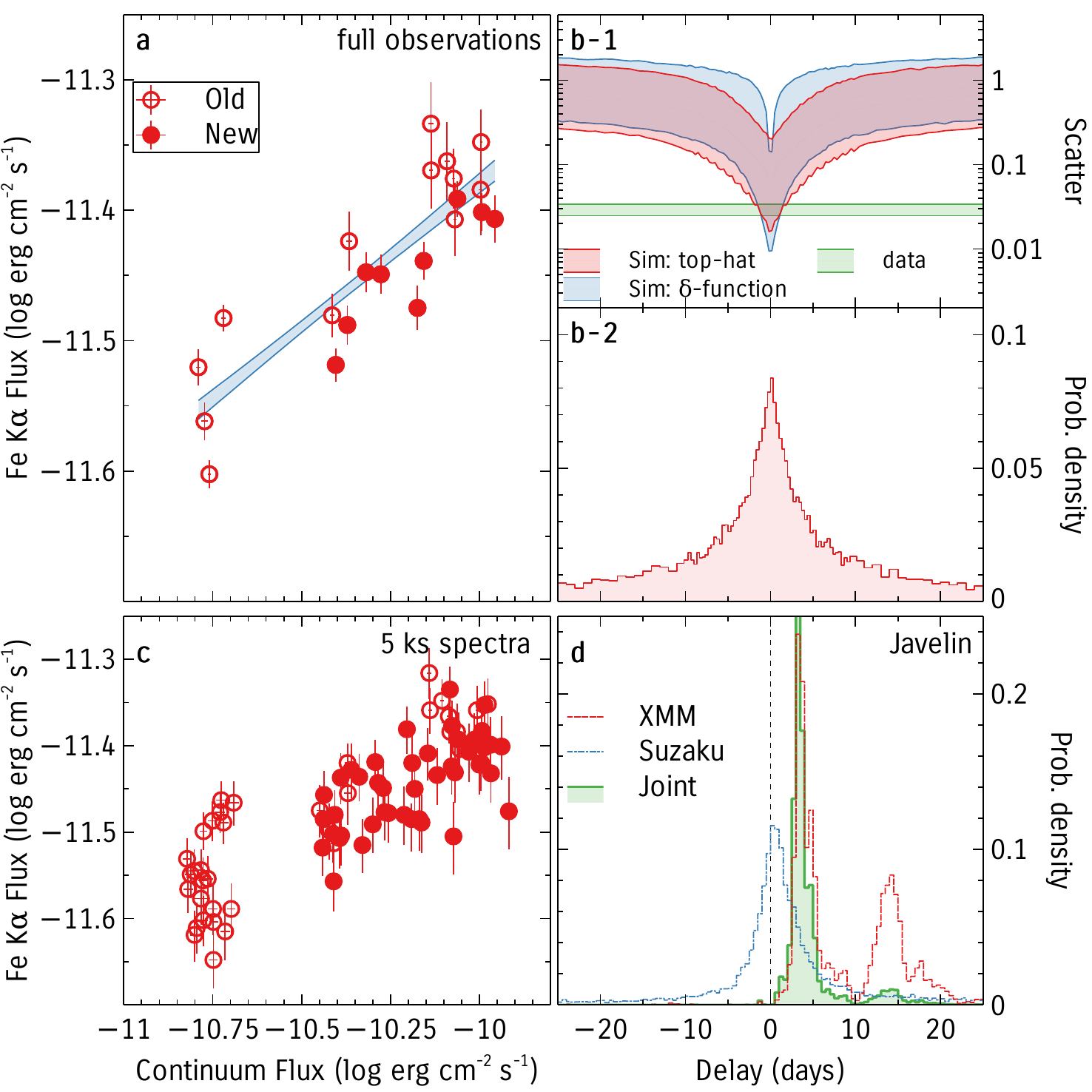}
\caption{Variability of narrow \FeK line in \ngcfour. {\bf a:} Variations of the \FeK line flux compared to the observed continuum flux. Old (observations 1--16) and new (observations 17--24) data are shown separately. The blue band shows the best fit linear model ($slope = 0.219\pm0.015$, $intercept = -9.19\pm0.16$). {\bf b-1:} Comparing the observed scatter in the line-continuum relation (green) from simulations assuming different delay values between the line and continuum variations. Simulations assuming a delta function and top hat transfer functions that explore possible geometries of the system are shown in blue and red respectively. {\bf b-2:} The probability density of the inferred delay between the line and continuum using the scatter distributions shown in b-1. {\bf c:} Similar to (a), but shown for spectra produced by splitting each observation to 5 ks segments. {\bf d:} Probability density of the time delay between the narrow \FeK line and the continuum resulting from modeling the 5 ks spectra with {\sc javelin}. \label{fig:cont_line}}
\end{figure}

We can estimate the time delay from the continuum-line correlation by noting that the scatter in the relation depends on the intrinsic variability of the continuum (i.e. its power spectra density; PSD), the sampling of the observations, and the time delay. For a given continuum PSD, if the observations are separated by more than the average delay, the continuum and the line will vary `in phase', and the scatter in their correlation will be minimal. If on the other hand, the sampling time is smaller than the time delay, the scatter is large, driven by the intrinsic random variability in the continuum. We measure the scatter in the observed data by fitting a linear model to the data shown in Figure \ref{fig:cont_line}-a (i.e. fluxes in log units), and take the square root of the sum of residuals as a measure of the scatter. For the observed data, we obtain a value of $(2.9\pm0.5)\times10^{-2}$.

To account for the randomness in the variability, we simulate light curves of the continuum and the line similar to those observed. We first estimate the intrinsic \emph{long term} power spectrum using the maximum likelihood method in \cite{2013ApJ...777...24Z}. We assume the power spectrum is a bending power law \citep{2004MNRAS.348..783M}. We estimate the PSD between $10^{-5}-5\, {\rm days}^{-1}$ to have an index of $-2.5\pm1.0$, a (natural log) normalization of $-9.3\pm0.6$ in rms units, and a (natural log) break frequency of $-2.9\pm1.3$ days$^{-1}$. These estimates are consistent with those measured using long term monitoring with RXTE \citep{2003ApJ...593...96M}. The estimated power spectrum from the line variations have is statistically consistent with that of the continuum. We use these parameters to simulate long term light curves of both the continuum and the line, with sampling similar to the observed, where the line is delayed with respect to the continuum with a delay between -30 and 30 days, using a simple delta-function transfer function. For each simulated set, we fit a linear model in the same way we did with the observed data and measure the scatter. This gives, for every lag value, a distribution of scatter values as different realizations of the light curve are produced. The resulting distribution of scatter values as a function of the assumed lag are shown in Figure \ref{fig:cont_line}-b1. 

The blue band from the simulation represents the 68\% confidence spread, and shows the expected trend for a fixed sampling pattern. As the delay increases (positive or negative), the scatter increases. The width of the band is controlled by the intrinsic random variability in the continuum light curve. Note that the scatter does not differentiate between positive and negative lags, hence the symmetry around zero lag. 

\begin{figure*}
\centering
\includegraphics[width=0.8\textwidth]{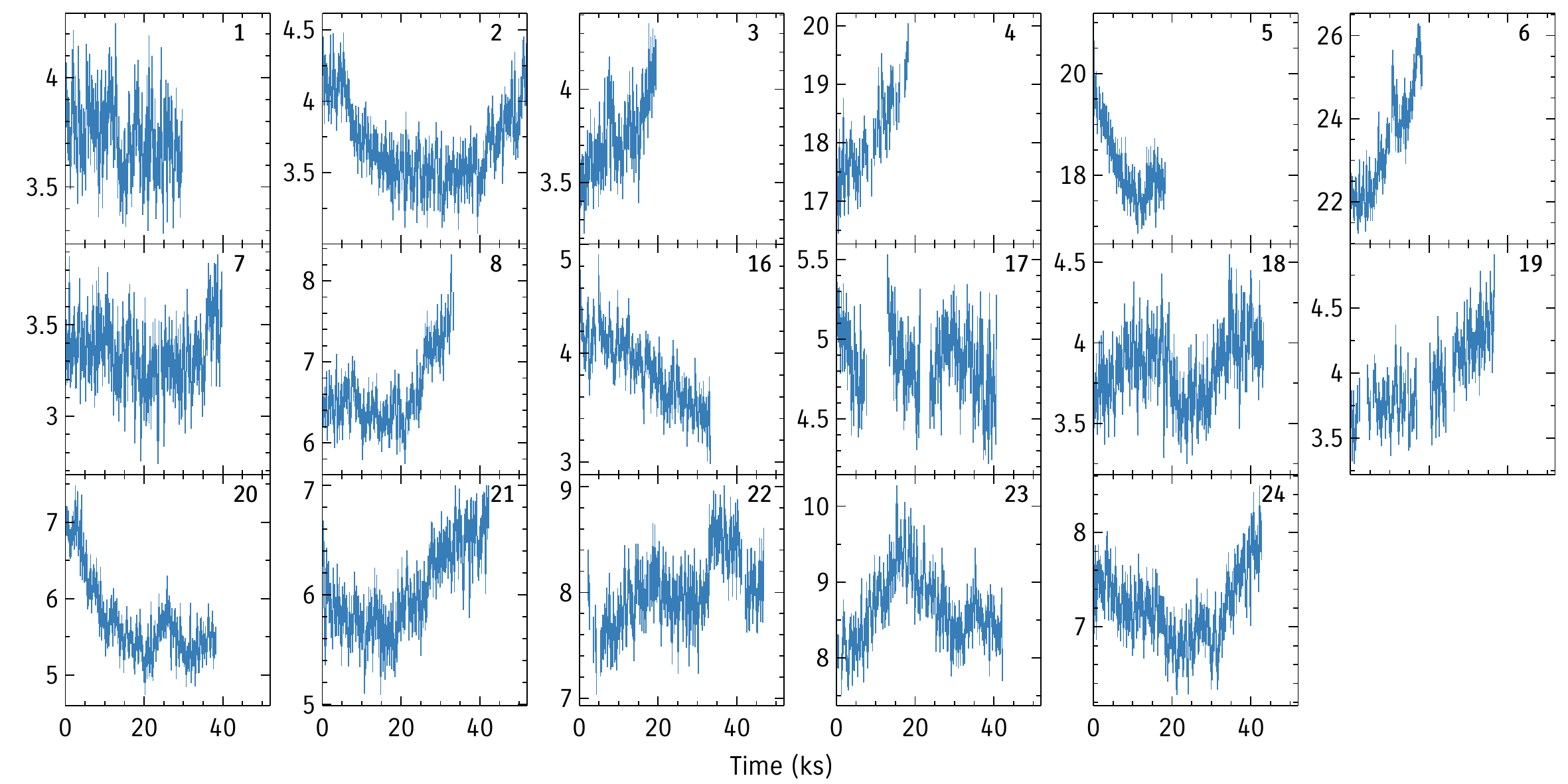}
\caption{2--10 keV light curves from 17 observations used in the fast timing analysis. These are identified by $(^t)$ in Table \ref{tab:obs_log}.}\label{fig:lc}
\end{figure*}

The plot in Figure \ref{fig:cont_line}-b1 shows that the measured scatter implies that the delay between the continuum and the narrow \FeK line is consistent with 0 days. The observed scatter is smaller than any scatter expected if the delay is larger than a few days. The probability density of the delay between the line and the continuum is shown in panel b-2.  The formal constraint we obtain for the delay is $0\pm2.8$ days ($1\sigma$). The probability density has broad side wings, with the 95\% confidence limit extending to 19 days. 

To explore the uncertainty in this modeling, we first use a power law PSD instead of a bending power law. We find that the scatter is always higher than shown in Figure \ref{fig:cont_line} because the power law PSD has less variability on time scales of 10--100 days than the bending power law model, and hence predicts less scatter. Second, we also explored the effect of the assumed delta-function response function between the continuum and the line. We repeat the simulations using a bending power law PSD and a top-hat response function whose width is drawn randomly between 0 and 10 days for every simulation. This is equivalent to a Bayesian assumption of a uniform prior on the width, and then marginalizing the final lag probability over this parameter. The results are shown with red band in Figure \ref{fig:cont_line}-b1. The broad response function reduces the variability in the line, producing less scatter. The probability density of the delay is not affected much, and it is shown in Figure \ref{fig:cont_line}-b2.

\subsubsection{Direct Estimate of the Narrow \FeK Delay}\label{sec:lag_javelin}
In order to estimate the lags directly, we split the observations to 5 ks segments ($\sim0.06$ days), and measure the delays using \textsc{javelin} \citep{2013ApJ...765..106Z}. The light curves are modeled with a Damped Random Walk process before estimating the delay. The correlation from these 5 ks segments are shown in Figure \ref{fig:cont_line}-c, and the resulting constraint on the delay is shown with the red probability density plot in Figure \ref{fig:cont_line}-d. 

The probability density has two peaks, the high peak at $\sim3$ days, and a smaller peak at $\sim14$ days. 
The shorter sampling provides by the 5 ks spectra covers more relevant ($\sim1$ day) time scales than those from individual observations, and therefore provides tighter constraints on the delay. Additionally, we tested if the measured lag vary or the two peaks in the probability density depend on flux. We found no evidence for such variability. The high flux data (splitting at continuum flux value of $=10.5$) gives the same two peaks, while the low flux data alone gives a broader peak that incorporate both peaks from the combined data.

Figure \ref{fig:cont_line}-d also shows in blue color the probability density resulting from performing the same analysis (over the 0.3--10 keV band) on all available \suzaku observations (observations IDs: 701034010, 707024010, 906006010 and 906006020). 
Although \suzaku XIS have a slightly better energy resolution than \xmm EPIC, the fact that the line is resolved in both detectors\footnote{The lower limit on the line width is significantly higher than the detector resolution at 6 keV in both cases.} implies that we are measuring the same component.
The second peak in the \xmm data could be due to the sampling, while the first peak and the main peak in \suzaku are statistically consistent. It is possible that the lag depends on the flux state. We have a NICER program to obtain more targeted observations to address this question.

The probability density from the combination of the data sets is shown in green in Figure \ref{fig:cont_line}-d. The secondary peak at 14 days is now weaker. Our best estimate of the delay between the \FeK line and the primary continuum from this combined data sets is $3.3^{+1.8}_{-0.7}$ days.

This is the first time such a direct estimate for the delay between the X-ray continuum and the narrow \FeK line is made. For comparison, the delay in optical reverberation mapping between the H$\beta$ line and the continuum at 5100 \AA\, is $6.6^{+1.1}_{-0.8}$ days \citep{2006ApJ...651..775B}. The small \FeK lag implies that the line is emitted from a region at about a factor of 2 \emph{smaller} than the optical broad line region (BLR). The result from these \xmm and \suzaku data provides independent confirmation of those obtained recently by modeling the shape of the narrow \FeK line and how it varies between flux intervals using the \chandra/HETGS \citep{2018arXiv180807435M}.


\section{Timing Analysis}\label{sec:timing}

Throughout the timing analysis, we focus on the 2--10 keV band as the softer band shows a reduced variability due to the small contribution from the nuclear emission. We extract light curves with 128 seconds bins in the 2--10 keV band. The light curves from the all the 17 datasets (see extraction information in section \ref{sec:data}) are shown in Figure \ref{fig:lc}. The plotted light curves have been re-binned to have a sampling of 256 seconds for plot clarity.

\subsection{The Power Spectrum}\label{sec:psd}
To estimate the intra-observation power spectral density, we calculate the periodogram from all the data combined. The periodogram is the squared amplitude of the Discrete Fourier Transform of the individual light curve segments. We use the rms normalization, and bin the frequencies by a geometric factor of 1.12. To reduce the red noise leak, we taper the light curves with a Hanning function \citep{Bendat:2000:RDA:555747}. We then model the resulting periodogram with a power law using Whittle statistics \citep{2010MNRAS.402..307V} as implemented in \textsc{xspec}. The best fit power law model has an index $\alpha=2.55^{+0.34}_{-0.30}$ and an integrated rms between $3\times10^{-5}-5\times10^{-4}$ Hz of $3.0\pm0.3$ \%. The slope is consistent with the steep power spectra measured for a large sample of Seyfert galaxies \citep{2012A&A...544A..80G}, and also consistent with our estimate from the long term variability in section \ref{sec:lag_scatter}.

\begin{figure}[h]
\centering
\includegraphics[width=0.7\columnwidth]{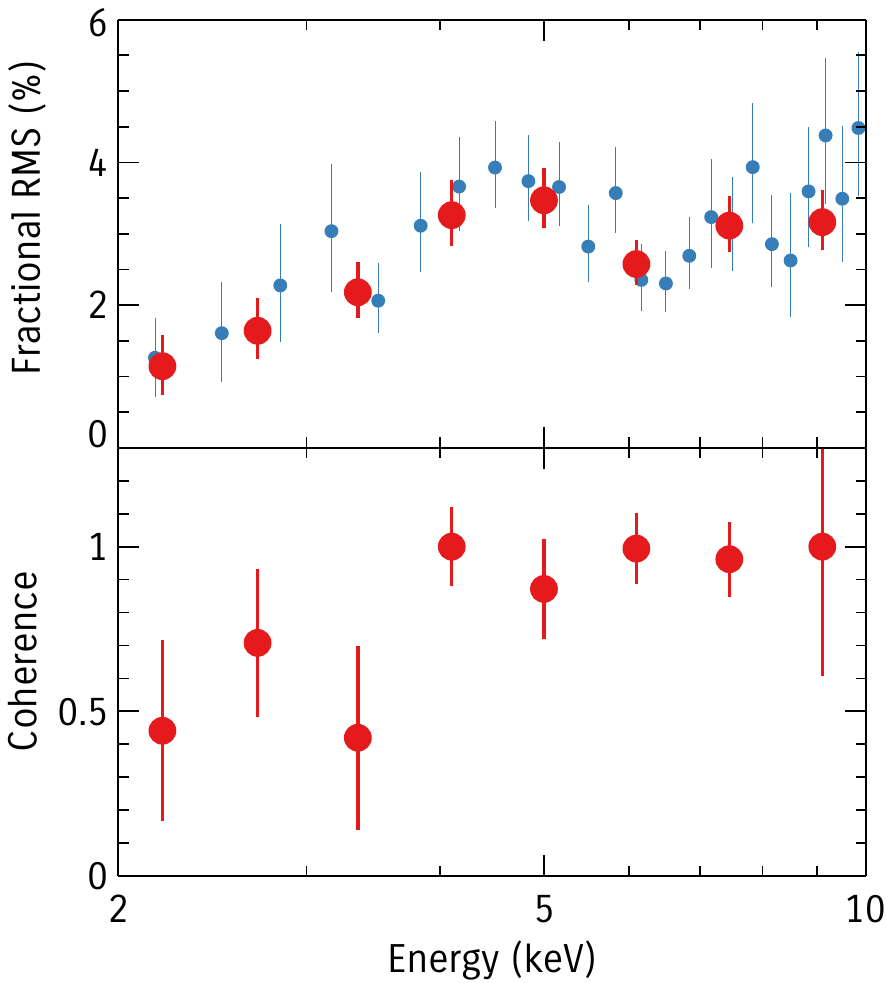}
\caption{{\em Top:} Fractional variability rms obtained by integrating the power spectrum between (0.3--5)$\times10^{-4}$ Hz. The red points are for 8 energy bins and the blue points are for using 24 logarithmically-spaced bins. {\em Bottom}: The coherence as a function of energy measured at the same frequency band.\label{fig:psd_coh}}
\end{figure}

\begin{figure}[t]
\centering
\includegraphics[width=0.7\columnwidth]{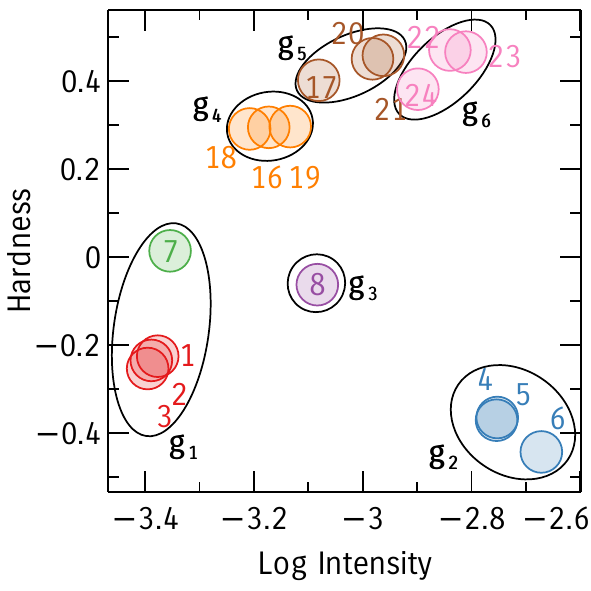}
\caption{Hardness-Intensity to show the observation groups used in Figure \ref{fig:lag_en}.\label{fig:hr}}
\end{figure}

We also calculate the periodogram at different energies. We use 8 bins between 2--10 keV in logarithmic steps. We follow a similar procedure to the total periodogram. We found that the data is consistent at the 99\% confidence level with a constant rms and slope. A significant change in the rms with energy is only seen if we assume a fixed slope, and the result of the integrated rms power in the (0.3--5)$\times10^{-4}$ Hz band is shown in the top of Figure \ref{fig:psd_coh}. The 8-bins data suggest that the variability power increases with energy with a bump at 4--5 keV. Doing the calculations for a larger number of bins (as shown in Figure \ref{fig:psd_coh}) suggests that the variability spectrum increases up to 4 keV and then flattens, with a drop at the energy of the narrow \FeK line. This shape resembles that of the absorbed primary continuum illustrated in Figure \ref{fig:show_spec}. The drop at low energy is due to the secondary component dominating below 4 keV. This conclusion is further illustrated by the bottom panel of Figure \ref{fig:psd_coh}, which shows the coherence \citep[e.g.][]{2014A&ARv..22...72U} as a function of energy for the same frequency band. The coherence high above 4 keV, suggesting a single component at those energies, and drops once the secondary power law component dominates.

\begin{figure*}
\centering
\includegraphics[width=0.85\textwidth]{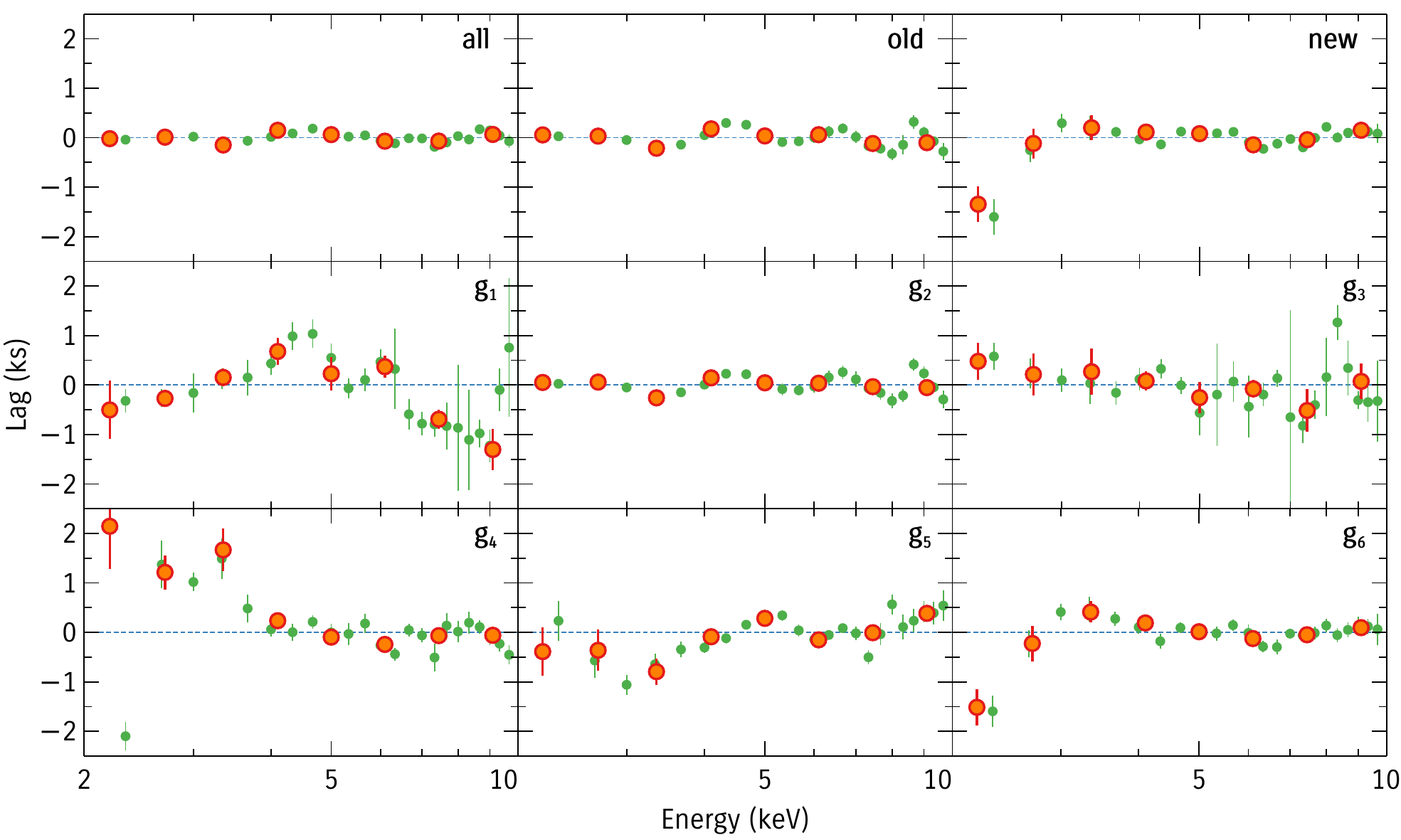}
\caption{Lag vs energy spectra from 17 \xmm observations of \ngcfour. The top three panels show the lags from all, the old (1--8) and new (16--24) datasets respectively. The lag measurements using 8 energy bins are shown as the orange circles with errors. The green points are produced by using 24 energy bins with logarithmic spacing, but instead of extracting the lag between every bin and the reference band, we average the lag ever every 2 neighboring bins, sliding the binning window by 1 every time. Every other point is therefore independent. The groups ${\tt g_1}-{\tt g_6}$ are defined in Figure \ref{fig:hr}. \label{fig:lag_en}}
\end{figure*}

\subsection{Time Delays}\label{sec:lag}
In \citetalias{2012MNRAS.422..129Z}, we presented detailed time delay measurements in the iron K band. The main result from that study was a detection of energy-dependent time delays, where the 5--6 keV band is delayed with respect to the other bands. The observations analyzed in that study are 1--8 as labeled in table \ref{tab:obs_log}. For consistency, we re-analyzed these observations along with the new (16--24) observations.

In \citetalias{2012MNRAS.422..129Z}, the lags were studied mainly in two frequency bands: $<2\times10^{-5}$ and $(0.5-5)\times10^{-4}$ Hz. The result in the first bin was mostly driven by the longest observation (obs-2 in table \ref{tab:obs_log}). All the new observations are shorter than obs-2, and do not provide any new information on that frequency band. We note that the number of variability cycles available at the lowest frequency band from obs-2 is small, and therefore the uncertainties in the lag measurements are likely underestimated \citep[e.g.][]{2016A&A...591A.113E}. Confirming the result in the lowest frequency band therefore remains to be tested with future observations.

We focus here on the higher frequency band (we use $(0.3-5)\times10^{-4}$ Hz to increase the number of variability cycles when considering observation groups). We extract light curves in 8 energy bins in the 2--10 keV band, with logarithmic binning. The light curves are extracted with a time sampling of 128 seconds. We calculate the inter-band time delays using the standard procedure \citep{1999ApJ...510..874N,2014A&ARv..22...72U}. The light curves are also tapered with a Hanning function similar to section \ref{sec:psd}. We measure the time delays as a function of energy using the whole 2--10 keV band as a reference, subtracting the light curve of the band of interest each time. 

We calculate lags from the combined dataset (labeled {\tt all}), and from individual subgroups. The {\tt old} and {\tt new} subgroups correspond to data from the old (1--8) and new (16--24) observations. We also calculate the lags for other subgroups that are produced using the K-means clustering algorithm to separate the observations into groups based on the hardness-intensity values. We used the counts in two bands, 2--3 and 8--10 keV, to calculate the hardness (H-S)/(H+S) and we use the hard band as measured of the un-absorbed intensity. A range of cluster numbers between 2--8 was explored. We find that 7 clusters gives a reasonable balance between the spectral complexity and signal in each group. The resulting hardness-intensity plot is shown in Figure \ref{fig:hr} with the each group plotted in a different color. The lag measurement from obs-7 was similar to obs1--3, so we grouped them into one group. The group labels (${\tt g_1}-{\tt g_6}$) and definitions are shown in Figure \ref{fig:hr}.

The resulting lag measurements are shown in Figure \ref{fig:lag_en}. The orange circles with errors are the lags from 8 energy bins. Figure \ref{fig:lag_en} also shows in green points a higher energy resolution version of the lag spectra. These are produced by using 24 energy bins with logarithmic spacing, but instead of extracting the lag between every bin and the reference band, we calculate the lags every 2 neighboring bins, sliding the binning window by 1 every time. Every other point is therefore independent. We show this to illustrate the effect of the energy bins and obtain a smoothed higher resolution version of the plots.

Panels ${\tt g_1-g_3}$ effectively reproduce the results in \citetalias{2012MNRAS.422..129Z}, where a lag spectrum that peaks at 5 keV was observed in the low flux observations (1--3, plus 7, which was not plotted in \citetalias{2012MNRAS.422..129Z}) but not in the high flux observations (4--6). Grouping the new observations by count rate is driven by the results in Z12 which shows that reverberation lags  are detected in the low flux observations and not in the high flux ones.

Panels labeled {\tt all} and {\tt old} are dominated by the highest flux observations (${\tt g_2}$). The spectra from the new data (${\tt g_4}$, ${\tt g_5}$ and ${\tt g_6}$) show a pattern that is different from the old data, and also different within the subgroups. The ${\tt g_6}$ panel is somewhat similar to ${\tt g_4}$, especially when we compare the higher resolution green points.

In order to assess the significance of these lags, we test the measurement against two null hypotheses: a constant lag and a log-linear change with energy. The first hypothesis is the simplest, and states that there is no inter-band time delays, and the lag-energy spectra should be a constant. The null model in this case has one free parameter. The second hypothesis allows for the possibility that there are continuum lags with a log-linear profile with energy without any spectral features. An increasing lag with energy is often observed to dominate at low frequencies, and may contribute to the frequencies studied here. The null model in this case has two parameters: a slope and an intercept, and the independent variable is the logarithm of the energy.

\begin{deluxetable}{l|c|c}
\tablecaption{Results of the hypothesis test for the lag spectra with 8 energy bins in figure \ref{fig:lag_en}. The table shows the confidence (in $\sigma$ units) at which the null hypothesis for a \texttt{constant} and a \texttt{log-linear} models is rejected. Values in bold are those with a rejection significance $>3\sigma$. \label{tab:lag_tests}}
\tablehead{Observations & 
{\tt constant} & 
{\tt log-linear} }
\startdata
{\tt all}	    & 2.9             & {\bf 3.2} \\
{\tt old}	    & {\bf 3.2}       & 2.3  \\
{\tt new}	    & {\bf 4.0}       & {\bf 7.6} \\
${\tt g_1}$		& {\bf 4.3}       & {\bf 6.7} \\
${\tt g_2}$		& 1.6       & 0.9 \\
${\tt g_3}$		& 0.4       & 0.4 \\
${\tt g_4}$		& {\bf 5.0}       & {\bf 6.0} \\
${\tt g_5}$		& {\bf 4.2}       & {\bf 3.1} \\
${\tt g_6}$		& {\bf 3.5}       & {\bf 5.9} \\
\enddata
\end{deluxetable}

The result of the tests are summarized in Table \ref{tab:lag_tests}. The table shows the confidence (in $\sigma$ units) at which the null hypothesis in each case is rejected. In other words, A value of 3, for example, means that the null hypothesis is rejected with $3\sigma$ confidence. The numbers in bold text highlight the cases where the null hypothesis is rejected at more than 3$\sigma$ level.

Table \ref{tab:lag_tests} shows that the ${\tt g1}$ group shows a significant energy dependent lag that is more complex than a constant and a linear trend. This quantifies the significance in the \citetalias{2012MNRAS.422..129Z} results. Inter-band lags are also significantly detected in the new observations, and the lag is more complex than linear in two groups in all subgroups. Note that the lags in ${\tt g4}$ {\em decrease} rather than increase with energy, as commonly observed for continuum lags.

The lags in ${\tt g_1}$ have been interpreted as being due to relativistic reflection. The peak at 5 keV suggests that the peak of a broad and redshifted iron line is delayed with respect to the continuum. The new data in groups ${\tt g_{4-6}}$ is different. We revisit the interpretation of these delays in section \ref{sec:discuss:broad}.


\section{Discussion}\label{sec:discuss}
The spectrum of \ngcfour is complex, with contributions from different emission and absorption components. We attempted in section \ref{sec:spec} to extract the main properties, and they are summarized in the following:

\subsection{Neutral Absorption}\label{sec:discuss:neutral_abs}
Neutral absorption is a significant contributor to the shape of the spectrum, as indicated by the strong Fe K edge at 7 keV and the curvature at 4 keV. The new data, taken within $\sim 40$ days, all have a column density of $\sim3\times10^{23}$ cm$^{-2}$, which is an order of magnitude higher than observed in previous years. Small changes in the column density are seen on time scales of days, but the largest changes are seen at time scales of weeks to months. 

The BLR, a likely origin for the absorbing material, has a scale size of several light days in \ngcfour \citep[][see also our discussion in section \ref{sec:discuss:narrow_line}]{2006ApJ...651..775B}. The changes of the absorption column on time scales of days is therefore likely associated with clouds in the inner BLR. The bulk of the column density changes, however, happen on longer time scales. If we model the power spectrum of the column density variations with a zero centered Lorenzian, we obtain a characteristic time scale of $t_{\rm abs}=36^{+39}_{-16}$ days. The width of the H$\beta$ line varies with time, with an average of $ 4000$ km s$^{-1}$ \citep{2006ApJ...651..775B}. The transverse velocity is $v = 4000 f$ km s$^{-1}$, where $f<2$ is a geometry factor whose upper limit can be obtained assuming Keplerian motion: $v_{\rm K} = (G M_{\rm BH}/R)^{1/2} < 8000$ km s$^{-1}$ (using $M_{\rm BH} = 3.6\times10^{7} M_{\odot}$). The radius $R$, is measured from the time delay of the narrow \FeK line; $R=3$ days (section \ref{sec:lag_javelin}), assuming it is produced in the same material. The measured time scale therefore implies a characteristic size for the absorbing structures of $s_{\rm abs}$ = $v t_{\rm abs} = f\times10^{15}$ cm ($=180f$ \rg, and for a scale, $R\sim1400$ \rg). Assuming these structures are spherical gives a density $n_{\rm abs} = N_{\rm H}/s_{\rm abs}$, and using the range of observed column densities $N_{\rm H} = 3-30\times10^{22}$ cm$^{-2}$, we obtain densities of $n_{\rm abs} = (3-30) f \times 10^{7}$ cm$^{-3}$.

The covering fraction of the neutral absorber that we used to model the extra emission at 2--4 keV appears to vary more than the column density itself. A characteristic time-scale estimate gives $t_{\rm c} = 4.8^{+6.0}_{-2.9}$ days. The parameters in table \ref{tab:fit_indiv} suggest that observations 17 and 18, for example, which are separated by 2 days, show variation about $7\%$, and although the exact number may depend on the details of the modeling and the parameters degeneracy, variations of at few percent level are observed between other observations too. If the extra component at 2--4 keV is due to a physical partial coverer, the small scale variations are expected, as the clouds move quickly across the line of the sight. This would imply that the $\sim 10^{15}$ cm structures are composed of smaller ($\sim 10^{13}$ cm) clouds that collectively produce the partial covering effect. This may also imply that the smaller clouds can have higher density than $10^{7}$ cm$^{-3}$ estimated for the larger structures, explaining why the clouds remain neutral despite being only a few light days from the central X-ray source.

If the 2--4 keV excess is modeled as a separate component, a power law for instance, the characteristic time scale increases only by small amount to $t_{\rm po} = 6.6_{-3.9}^{+8.3}$ days, and it remains smaller than the characteristic time scale of the absorber changes. This implies that this additional component originates in a region smaller than the BLR producing the neutral absorption, or that its flux is affected by structures close to the line of sight.

\subsection{Warm Absorption}\label{sec:discuss:wa}
For the low $\xi$ warm absorber, there are indications in the data that it is more ionized when the flux is high. Two observations do not seem to fit the trend, 1 and 2, which have low fluxes with ionization parameter $log\xi=1.3-1.5$. Scanning the $\xi$ parameter space, we find that in both cases, there are other solutions with only marginal difference in $\chi^2$ statistic where the ionization is low ($\Delta\chi^2=10, 12$ for observations 1 and 2 respectively). These low ionization solutions appear to be more physical and in line with other observations. Forcing the ionization to be low for observations 1 and 2, we find a correlation between the hard (7--10 keV) flux and the ionization of the warm absorber (Spearman rank correlation coefficient $r=0.51$, $p=7.5\times10^{-3}$), with a best fit linear model (in log units) of $log \xi = (1.9\pm0.2) \times log F_{\rm hard} + (20\pm2)$. We consider the correlation as suggestive given the model degeneracy at the lowest flux observations (observations 1 and 2), and it would imply that the warm absorber is responding to changes of the continuum. The index of the correlation, however, is not equal to unity, implying that photoionization is not the only source of variability and changes in the density and/or location of the absorber are needed. It is conceivable that layers of the same neutral absorber in the BLR are being ionized by the increasing flux of the source. Future observation, at low fluxes specifically, may allow the location of the absorber to be independently measured by tracking these ionization changes.

The data also suggest that the high ionization absorber is also more prominent at higher flux observations. This apparent by the fact that the lowest flux observations (1, 2 and 3) show no signs of this component. The absorber is consistent with zero velocity (an upper limit of $400$ km s$^{-1}$), similar to the low ionization component, suggesting that they are part of the same absorbing system.

\subsection{Narrow \FeK Line}\label{sec:discuss:narrow_line}
The narrow \FeK generally shows little or no variability within a typical exposure (less than 1 day) time scale, while it varies on time scales of years, at least in some sources. Long term monitoring with RXTE suggested that only a small fraction of observed sources show variability in the line \cite{2003ApJ...598..935M}. Four out of 15 sources studied using ASCA showed changes in line intensity, with \ngcfour showing no variability \citep{2001ApJ...550..261W}. On much longer time scales (1000--2000 days) that correspond to light travel sizes larger than the standard torus, variability in the line is common \citep{2016ApJ...821...15F}.

Changes in the narrow \FeK line flux or ionization in individual observations separated by a few to several hours have been reported only for a few sources, including Mrk 841 \citep{2002A&A...388L...5P}, NGC 7314 \citep{2003ApJ...596...85Y}, Mrk 509 \citep{2013A&A...549A..72P}, NGC 2110 \citep{2015MNRAS.447..160M} and NGC 2992 \citep{2018MNRAS.478.5638M}.

Evidence for a direct link between the variability of the narrow \FeK line and the continuum were observed NGC 4051 from RXTE monitoring \citep{2003MNRAS.338..323L}. More recently, an intesive campaign on Mrk 509 \citep{2013A&A...549A..72P} with \xmm revealed that the line intensity is correlated with continuum flux. Similar trends with smaller number of observations were also observed in NGC 2110 and NGC 2992 \citep{2015MNRAS.447..160M,2018MNRAS.478.5638M}. Our results confirm a similar relation in \ngcfour, with many more observations and smaller observational uncertainties. 

It is generally thought that the dust sublimation radius forms an outer envelope to the \FeK line emitting region \citep{1993ApJ...404L..51N,2011A&A...525L...8C,2015ApJ...812..113G,2018MNRAS.474.1970B} 
Although \cite{2006MNRAS.368L..62N} found no correlation between the width of the \FeK and H$\beta$ line in a sample of sources, implying that the \FeK line originates in the torus and not the BLR, other observations suggest otherwise \citep[e.g][]{2008MNRAS.389L..52B}. \cite{2010ApJS..187..581S} presented an extensive analysis of the \chandra grating spectra of 36 sources, finding that there is no universal location for the \FeK line-emitting region relative to the optical BLR. They find that a given source may have contributions to the \FeK line flux from parsec-scale distances from the putative black hole, down to matter a factor $\sim 2$ closer to the black hole than the BLR. This latter result is consistent to our conclusion, based on the variability of the line.

The simplest interpretation of the shorter delay of the \FeK line compared to the H$\beta$ line is that it originates in a smaller region. Shorter lags can also be produced if the \FeK line originates in a structure that is closer to our line of sight than the optical BLR. The small delay in this case is a consequence of the shorter \emph{path difference} between continuum and line photons. One scenario where this is possible is that the \FeK line is produced in the outer surface of the torus that is grazed by our line of sight to the central source. The delays in this case are small even if the distance between continuum source and the torus is large. The geometry is similar to the scattering reverberation model proposed to explain the relativistic reverberation in AGN \citep{2010MNRAS.403..196M}. This scenario is however unlikely as it would imply that most of the gas emitting the line is located close to our line of sight. The scenario is also unlikely given the \chandra measurements of the \FeK line asymmetry and the implied region size, which is smaller than that inferred from the H$\beta$ line \citep{2018arXiv180807435M}. Therefore, we are left with the hypothesis that the \FeK line is emitted from a region smaller than the optical BLR. This could be the inner regions of the BLR itself, or the cold accretion disk. It is likely that the variable narrow \FeK line is produced in the same clouds responsible for the absorption discussed in section \ref{sec:discuss:neutral_abs}.

The time lag of the narrow \FeK line we find is comparable to the delay between the X-ray and the UV \citep{2017ApJ...840...41E}. The UV delay was longer than the light-travel delay predicted by models of simple reprocessing in a thin disk. A process with a characteristic time scale longer than the light-travel time was therefore invoked \citep{2017MNRAS.470.3591G}. If such a model is correct, the similarity of the two delays is just a coincidence. If the X-ray/UV delay is due to light travel time, however, then the sizes of the UV-emitting region and the \FeK line region are comparable in size, though not necessarily co-located.

\subsection{The Black Hole Mass from the Narrow \FeK Delay}\label{sec:discuss:mass}
Estimating black hole masses from optical reverberation measurements is well established \citep{2004ApJ...613..682P}. We can apply the same technique given the delay we obtained and a measure of the line width. The black hole virial mass is related to the observed quantities by:
\begin{displaymath}
M_{\rm BH} = \frac{fc\tau v^2}{G}
\end{displaymath}
where $\tau$ is the time delay, $v$ is the velocity of the line emitting gas, inferred from the line width. The factor $f$ again depends on the structure, kinematics, and orientation of the emitting region. 
Our measured line widths are consistent with those obtained from the higher resolution HETG \chandra data \citep{2018arXiv180807435M}, we therefore use their estimate of the line width as it has lower statistical uncertainty. The line width is $\sigma = 23\pm2$ eV, but for the purpose of the mass estimate, it is more reasonable to use the width of the \emph{variable} part of the line. Here, we use the line width measured in the difference spectrum, which is $55^{+42}_{-22}$ eV \citep{2018arXiv180807435M}.
Using the measured  lag of $\tau = 3.3^{+1.8}_{-0.7}$ days, we obtain a mass of
$M_{\rm BH} = (4.3^{+5.2}_{-2.6}) f \times10^{6} M_{\odot}$. In optical reverberation, $f$ is obtained by scaling masses to fall on the $M$-$\sigma$ scaling relation. The most recent estimates give a value of $f=4.13\pm1.05$ \citep{2013ApJ...773...90G}. Assuming the \FeK emission region has similar geometry to the BLR, the black hole mass becomes $M_{\rm BH} = (1.8^{+2.2}_{-1.1}) \times10^{7} M_{\odot}$. This is a factor of $\sim 2$ smaller than other black hole estimates, which are: $M_{\rm BH} = 3.57^{+0.45}_{-0.37}\times10^{7} M_{\odot}$ from optical reverberation mapping (\citealt{2006ApJ...651..775B}; but using more recent calibration scale by \citealt{2013ApJ...773...90G}), $M_{\rm BH} = 3.0^{+0.75}_{-2.2} \times10^{7} M_\odot$ from gas dynamics \citep{2008ApJS..174...31H}, and $M_{\rm BH} = (3.76\pm1.15)\times10^{7} M_{\odot}$ from stellar dynamics. The relatively large uncertainties, however, make our measurements formally consistent with the other estimates. 
It is also possible that a larger $f$ is needed, and hence a different geometry for the \FeK-emitting region compared to the BLR. More precise measurement of the width of the variable part of the line in the future (e.g. with \chandra, XRISM or Athena) will be able to reduce this uncertainty.

\subsection{Relativistic Reflection \& Reverberation}\label{sec:discuss:broad}
We showed in section \ref{sec:spec:toward_mod} that the spectrum of \ngcfour is complex, requiring non-uniform neutral absorption and at least two layers of warm absorption. After accounting for these features, we find no strong evidence or requirement for a relativistic broad line. This is particularly the case for the new data, and by extending the best fit model to the old data, the same statement can be made. An upper limit to the contribution of relativistic reflection is around 3\% in individual observations, and about 1\% in the combined spectrum.

We discussed in section \ref{sec:full_model} why published evidence for a relativistic line \ngcfour needed to be revisited (there are other similar cases in the literature; for example, Fairall 9; \citealt{2016MNRAS.462.4038Y}). It remains the case, however, that the strongest evidence was the detection of reverberation. Variations in light curves at $\sim 5$ keV are observed to lag those at lower and higher energies. The shape of the lag as a function of energy traces the shape of broad relativistic line (\citetalias{2012MNRAS.422..129Z}; see panel \gone in Figure \ref{fig:lag_en}), suggesting the presence of broad relativistic line in the spectrum. We find that the shape of the lag-energy spectrum in the new data is {\em different}, as detailed in Figure \ref{fig:lag_en}.

Interpreting the lag in the old data was relatively straight forward. The difference between ${\tt g_1}$, ${\tt g_2}$ and ${\tt g_3}$ was attributed to the level at which relativistic reflection contributed to the overall spectrum. In ${\tt g_1}$, it contributes significantly and therefore leads to observable inter-band delays and a peak at 5 keV. In ${\tt g_2}$ and ${\tt g_3}$, its contribution is minimal, and no lags are observed. In the new data, lags are present, but {\em they do not resemble a relativistic Fe line shape}.

The spectra of the new data are dominated between 2--4 keV by the extra component that can be due to either scattering or intrinsic emission leaking through the absorber (Figure \ref{fig:apdx:fit4_all}; see also the spectra averaged over the groups used in the lag analysis in Figure \ref{fig:spec_group}). It is tempting therefore to associate the observed lags with this component. \cite{2010MNRAS.408.1928M} suggested that time delays seen in many sources are due to reverberation of a large reflecting medium. The energy dependence of the lag results from the energy-dependent reflection fraction, which generally has a hard shape. It is not unreasonable that the interplay between an absorbed primary continuum and the distant reflection spectrum may produce complex shapes similar to those observed. This model, however, predicts strong narrow features at the 6.4 keV produced by the strong narrow line present in the reflection spectrum. The lags in Figure \ref{fig:lag_en} suggest a smooth curve around 6.4 keV for most cases.

The variability in the 2--4 keV band  can also be affected by changes in the absorbing system, such as the column densities and covering fraction of the neutral absorber (or equivalently, the flux of scattering component), and the column density or ionization of the warm absorber. A variation of any of these parameters, along with the intrinsic continuum, may produce complex lag-energy patterns. This is not because such parameters are changing in response to the continuum, but rather, it is because a smooth change in these parameters affects the spectrum in a non-linear way, making some energy bands appear to respond before others.

\begin{figure}
\centering
\includegraphics[width=\columnwidth]{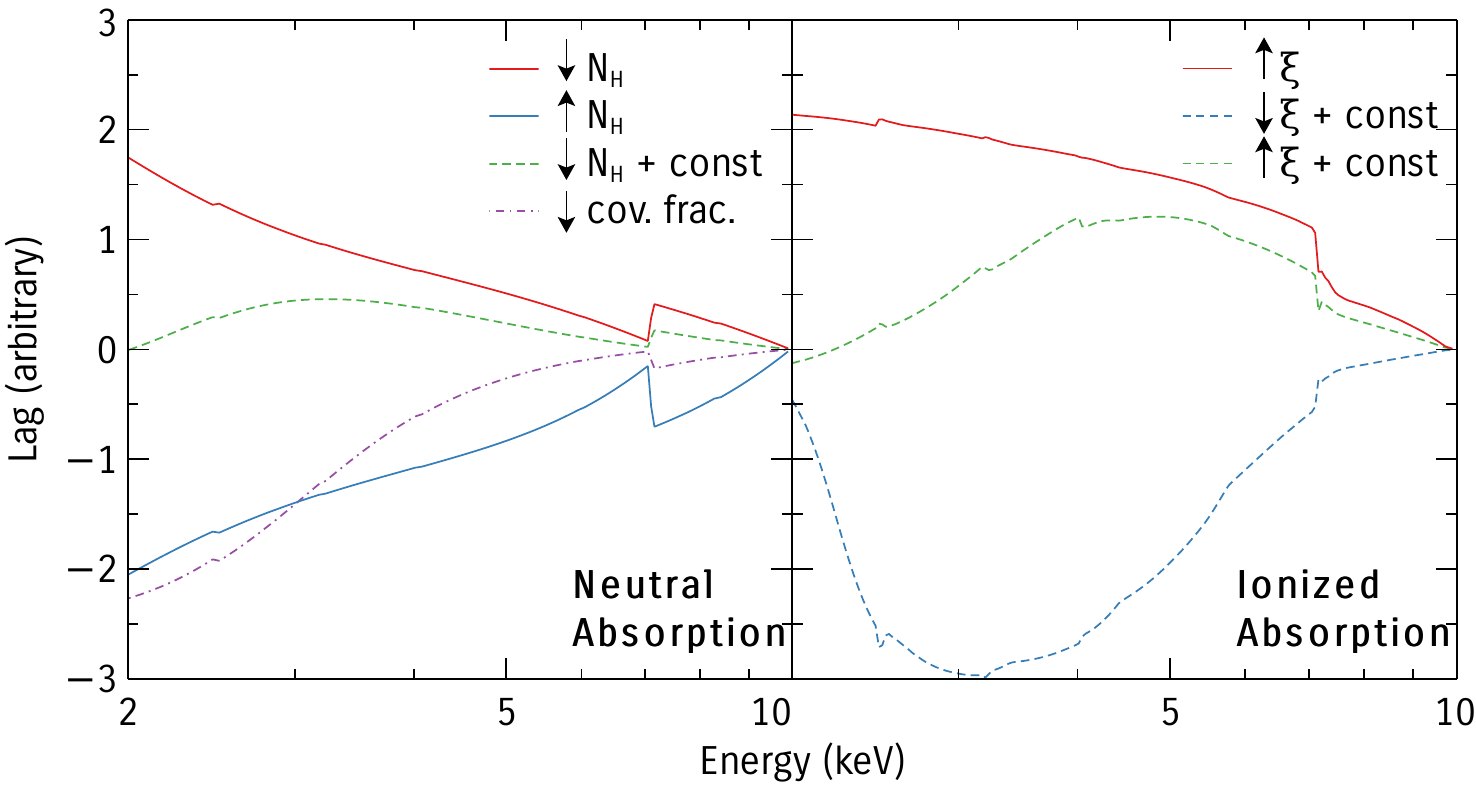}
\caption{Simple lag models assuming variability in the absorption. The lags are calculated relative to the last bin, and the vertical shift, as well as the actual lag values are arbitrary. The left panel is for varying the column density $N_{\rm H}$ and the covering fraction of a neutral absorber. The right panel is for varying the ionization parameter $\xi$ for ionized absorption. Note that adding a non-variable component that dominates between 2--3 keV, dilutes the lags, producing a peak or trough at $\sim 5$ keV.\label{fig:lag_model}}
\end{figure}

To illustrate this effect, we consider a neutral absorber in front a source emitting a power law spectrum. For simplicity, we can imagine two sine wave at two energy bands representing the intrinsic source variability. When the column density is constant with time, the two waves rise and fall at the same time, giving a zero delay between the two bands. If, on the other hand, the column density changes, the rise and fall time time of the waves will be different for the soft and hard band owing to the $e^{-N_{\rm H}(E, t)\sigma(E)}$ factor from the absorption that multiplies the spectra, where $\sigma$ is the absorption cross section. A similar effect can also result from a varying covering fraction in a non-uniform absorber, or the ionization parameter in a warm absorber. 

The non-linear dependence of the spectra on these parameters produces complex delay patterns, some of which are illustrated in Figure \ref{fig:lag_model} \citep[see also][]{2016A&A...596A..79S}. We show examples of lag-energy models that result from varying some of the absorption parameters for neutral (left) and ionized (right) absorption. An earlier example of these models was presented in \cite{2015MNRAS.446..737K}. We assume the intrinsic continuum is a power law with a photon index of 1.7 and no intrinsic delay between different energies. We then assume the parameters of interest (e.g. $N_{\rm H}$, $\xi$) vary linearly by 20--30\%. In these simplified models, we assume the driving variability is a sine wave with 10\% amplitude variations. Spectra at 1000 time steps are generated from the variable absorption model in which $N_{\rm H}$, $log \xi$ or the covering fraction vary linearly between 5--15, -1--0 and 0.5--0.6 respectively. The resulting variable spectra are used to calculate the lag as function of energy.

In the left panel, we see that a decreasing or increasing column density produces a decreasing or increasing lag spectrum, with an apparent Fe K edge feature. We also experimented with adding a non-variable component that contributes significantly at 2 keV. This illustrates the effect of a non-variable (at the time scales probed) distant reflection or scattering component. Its effect is to dilute the lags toward 2 keV (green dashed curve).

The effects of the ionization parameter $\xi$ are illustrated in the right panel, where in this case we use {\tt zxipcf}, fixing the column density at 10 and assuming the $log \xi$ changes between -1 and 0. The increasing ionization allows more soft flux to penetrate the absorber, and produces a curve similar to decreasing column density in the neutral case. When a constant component is added, the lags are again diluted producing a peak or trough at $\sim 5$ keV, depending on whether $\xi$ increases or decreases. The energy of the peak/trough depends on the column density of the absorber. The parameters in Figure \ref{fig:lag_model} were selected to produce lag spectra that broadly resembles those we observed.

Figure \ref{fig:lag_model} shows that the complexity of the lags observed in \ngcfour is likely due to variability in the absorption system. Although we have not formally fitted these models to the lag spectra, our analysis of the 5 ks spectra (used for the measuring the narrow \FeK lag in section \ref{sec:narrow_lag}) indicated that changes in the absorption parameters are common. This is in line with previous work using principle component analysis \citep{2015MNRAS.447...72P}. The same work also finds that other sources show variability in a relativistic component. Figure \ref{fig:lag_model} suggests that the presence of variable absorption in the line of sight may complicate the interpretation of the energy-dependent time lags. Our current work demonstrates this complexity for the case of \ngcfour. NGC 1365 appears to share many properties of \ngcfour \citep{2015MNRAS.447...72P}, so the interpretation of the spectra and the spectra may be related. Other sources, where absorption is not as prominent, may need a case-by-case analysis. For instance, hard X-ray lags are seen in `bare` AGN such as NGC 6814 \citep{2013ApJ...777L..23W}.

\section{Summary}
The main results can be summarized in the following:
\begin{enumerate}

\item Studying the spectrum of the brightest Seyfert nucleus over two decades reveals that neutral and ionized absorption play a significant role in shaping the spectrum.

\item The data reveals a highly variable narrow \FeK line on time scales of days. We measure for the first time, a time delay between the narrow \FeK line and the continuum of $\tau = 3.3^{+1.8}_{-0.7}$ days, implying that the line is produced in the inner BLR. Assuming the X-ray emitting BLR has a similar geometry to the optical BLR imply a black hole that is half that obtained from the optical reverberation. If we assume, on the other hand, that the mass from the optical reverberation mapping is correct, then the X-ray emitting BLR appears to have a {\em flatter} geometry.

\item We observed weak but significant days time scale variability in the soft band ($<1$ keV) thought to be dominated a distant  emission spectrum. The variability implies that either some nuclear emission leaks through the absorbers and contributes to the soft band and/or the photo-ionized gas is located at only a few light days from the central source.

\item Accounting for the absorbers in the system, we find no strong evidence or requirement for relativistic reflection from the inner disk.

\item Significant energy-dependent time lags are measured. The lag-spectra in the new do not resemble a simple broad iron line, unlike the old data. The new lag spectra suggest that the lags are not due to reverberation of the relativistic component, but rather due to variations in the absorbing system.

\end{enumerate}

\acknowledgments
This material is based upon work supported by NASA under Grant No. NNX15AW08G issued through the XMM-Newton GO program. This work is based on observations obtained with XMM-Newton, an ESA science mission with instruments and contributions directly funded by ESA Member States and NASA.

\appendix
\section{Additional Modeling Details}\label{apdx:fit4_all}
\begin{figure}[t]
\centering
\includegraphics[width=\textwidth]{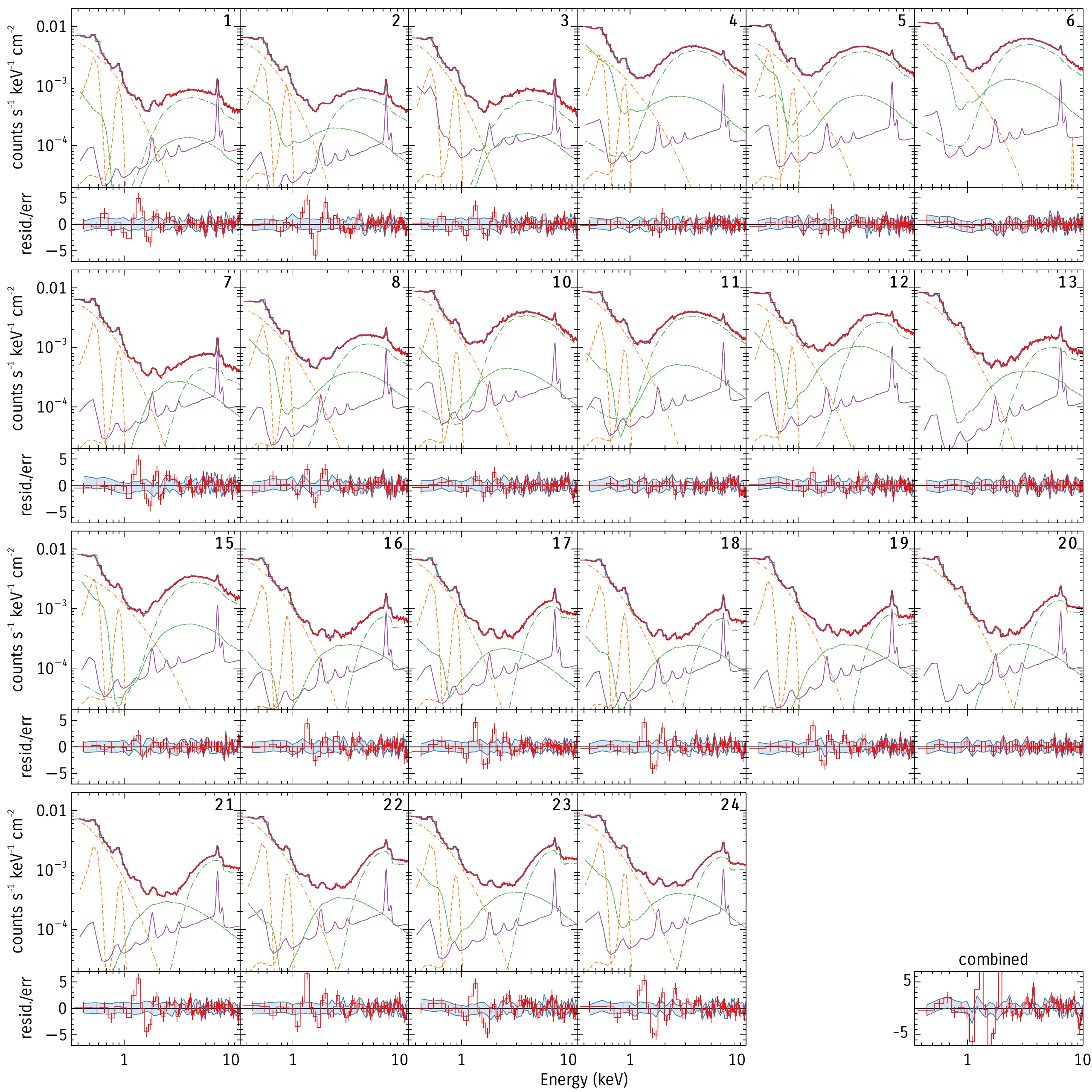}
\caption{Results of model fitting from section \ref{sec:full_model}. This is similar to Figure \ref{fig:fit4_4spec} in the main text but shows all the spectra. {\em Spectral row}: The data is shown in red, the total model in blue. The primary (absorbed) power law is shown in dot-dashed green and the secondary is shown in dashed green. The distant reflection component is in purple, while the soft components are shown in orange. {\em Residuals row}: The red points are the residuals for the model that include only the strongest soft lines. The blue bands are the residuals when additional emission lines are added to the soft band (See text for details).\label{fig:apdx:fit4_all}}
\end{figure}

\begin{figure}[t]
\centering
\includegraphics[width=\textwidth]{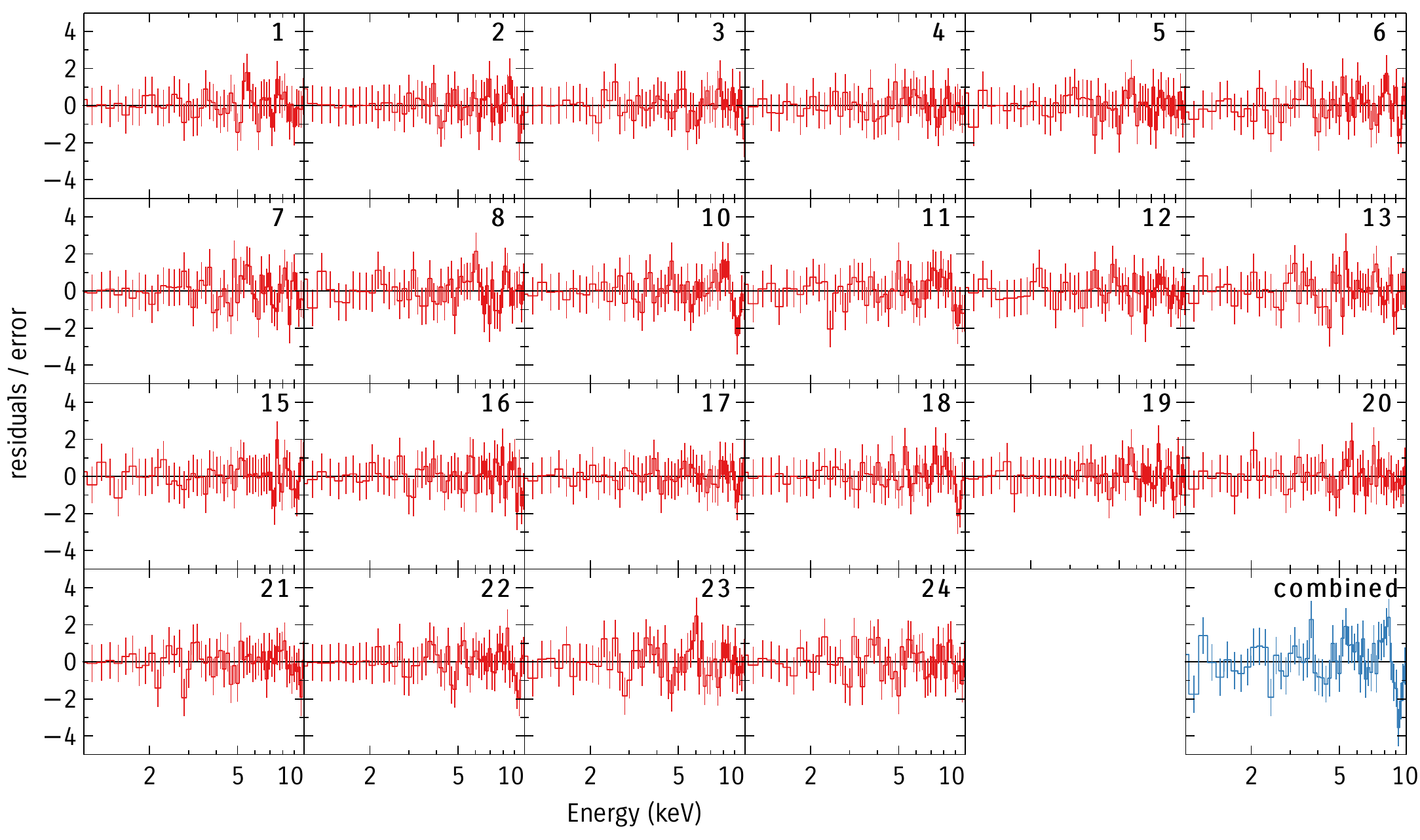}
\caption{Residuals to the spectral model after taking emission lines in the soft band into account. Weak absorption lines at 2.3, 2.9, 4.9 and 9.2 keV are observed in several observations, and in the combined residuals. The lines are weak in individual observations, but the fact some are seen in multiple observations suggests they are real.\label{fig:apdx:fit4_4a}}
\end{figure}

\begin{figure}
\centering
\includegraphics[width=0.6\textwidth]{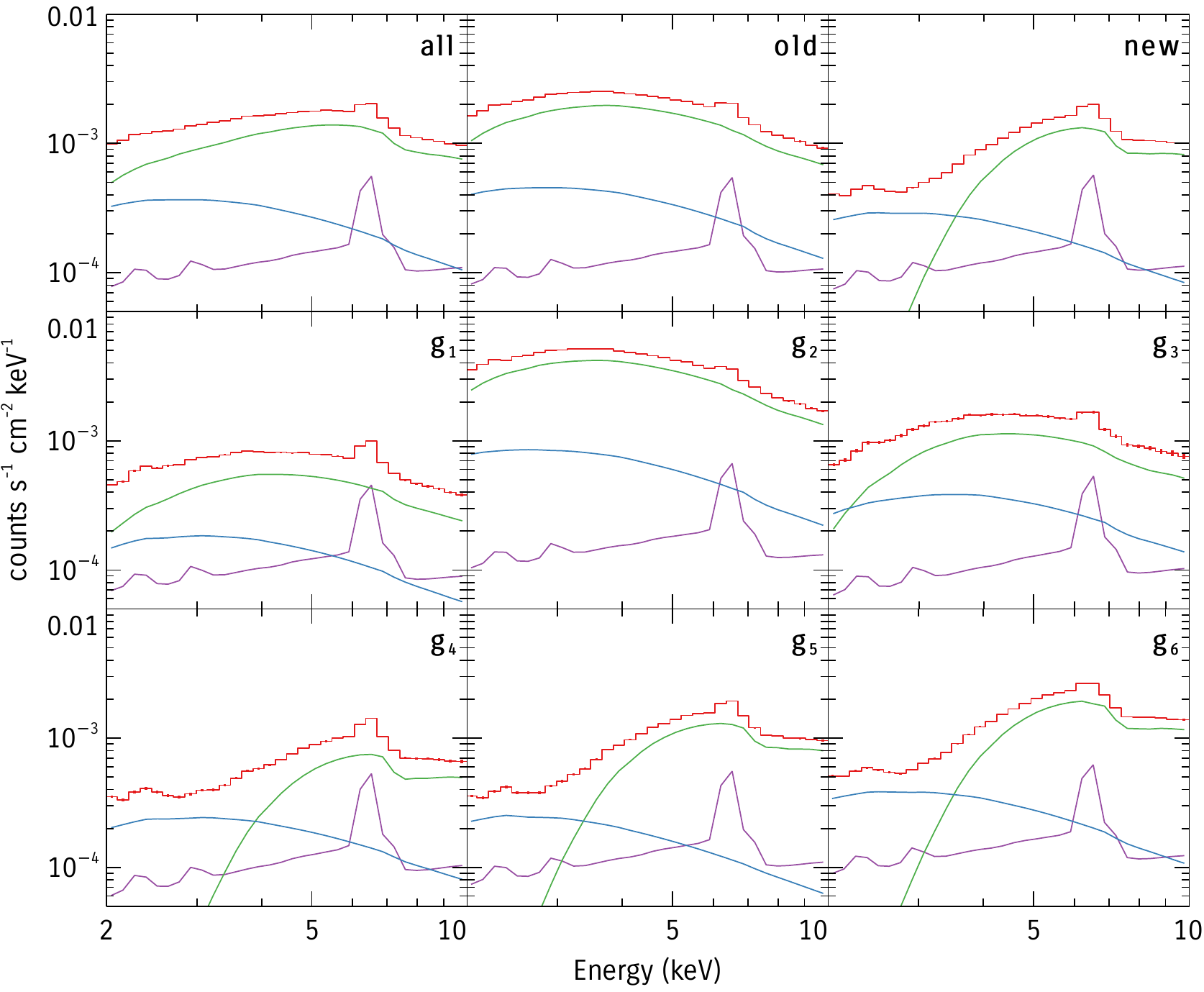}
\caption{The spectra and best model averaged over the groups used in the timing analysis, defined in Figure \ref{fig:hr}. The energy axis has been also re-binned to 32 bins to close to those used in the lag plot in Figure \ref{fig:lag_en}.\label{fig:spec_group}}
\end{figure}

\begin{longrotatetable}
\begin{deluxetable}{lccccccccccccc}
\tabletypesize{\scriptsize}
\tablecaption{Parameters of the best fit model from individual spectra.\label{tab:fit_indiv}}
\tablehead{
\colhead{Component} & \colhead{\xsm{zxipcf}{l}} & \colhead{\xsm{zxipcf}{l}} & \colhead{\xsm{zxipcf}{h}} & \colhead{\xsm{zxipcf}{h}} & 
    \colhead{\tt zTBabs} & \colhead{\xsm{po}{h}} & \colhead{\xsm{po}{h}}& \colhead{\xsm{TBpcf}{}} & 
    \colhead{\tt xillver} & \colhead{\xsm{brems}{s}} & \colhead{\xsm{brems}{s}}   & \colhead{} & \colhead{}
\\[-0.2cm]
\colhead{Obs} & \colhead{$N_{\rm H}$} & \colhead{$log\xi$} & \colhead{$N_{\rm H}$} & \colhead{$log \xi$} & 
    \colhead{$N_{\rm H}$} & \colhead{F (7--10)} & \colhead{$\Gamma$} & \colhead{$c_f$} & \colhead{F (6--6.8)} & \colhead{$kT$} & \colhead{$N_{\rm s}$} & \colhead{$\chi^2/$d.o.f} & \colhead{$p_{\rm null}$}
    \\[-0.2cm]
\colhead{} & \colhead{$10^{22}$ cm$^{-2}$} & \colhead{} & \colhead{$10^{22}$ cm$^{-2}$} & \colhead{} & 
    \colhead{$10^{22}$ cm$^{-2}$} & \colhead{log \fluxu} & \colhead{}& \colhead{} & 
    \colhead{log \fluxu} & \colhead{keV} & \colhead{$10^{-3}$}   & \colhead{} & \colhead{}
    }
\startdata
1   & $8.9\pm1.7$           & $1.6\pm0.1$           & -                     & -                     & 
    $1.1\pm0.9$             & $-11.14\pm0.01$       & $1.55\pm0.06$         & $0.84\pm0.07$         &
    $-11.56\pm0.01$         & $0.57\pm0.03$         & $3.7\pm0.4$           & $92/82$   & 0.21 \\
2   & $4.7\pm0.8$           & $1.4\pm0.1$           & -                     & -                     & 
    $3.8\pm0.9$             & $-10.762\pm0.003$     & $1.48\pm0.02$         & $0.84\pm0.02$         &
    $-11.60\pm0.01$         & $0.41\pm0.01$         & $4.5\pm0.2$           & $80/76$   & 0.34 \\
3   & $5.5\pm0.3$           & $-0.4\pm0.1$          & -                     & -                     & 
    $<1.3$                  & $-10.792\pm0.008$     & $1.87\pm0.04$         & $0.80\pm0.13$         &
    $-11.52\pm0.02$         & $0.41\pm0.04$         & $4.4\pm0.4$           & $86/72$   & 0.13 \\
4   & $6.1\pm1.0$           & $1.5\pm0.2$           & $1.5\pm0.7$           & $3.4\pm0.1$           & 
    $2.2\pm0.6$             & $-10.071\pm0.004$     & $1.66\pm0.03$         & $0.89\pm0.03$         &
    $-11.40\pm0.03$         & $0.95\pm0.07$         & $2.5\pm0.4$           & $74/85$   & 0.87 \\
5   & $7.6\pm0.2$           & $1.4\pm0.1$           & $18\pm12$             & $4.8\pm0.5$           & 
    $0.4\pm0.1$             & $-10.073\pm0.005$     & $1.63\pm0.03$         & $0.85\pm0.02$         &
    $-11.37\pm0.03$         & $0.7\pm0.2$           & $3.5\pm0.6$           & $99/91$   & 0.26 \\
6   & $4.7\pm0.3$           & $1.2\pm0.1$           & $2\pm0.6$             & $3.3\pm0.1$           & 
    $1.9\pm0.2$             & $-9.996\pm0.002$      & $1.69\pm0.02$         & $0.84\pm0.04$         &
    $-11.34\pm0.01$         & $0.8\pm0.1$           & $3.5\pm0.5$           & $112/88$  & 0.04 \\
7   & $3.0\pm0.6$           & $-1.7\pm0.4$          & -                     & -                     & 
    $1.4\pm0.7$             & $-10.72\pm0.01$       & $1.61\pm0.03$         & $0.80\pm0.01$         &
    $-11.48\pm0.01$         & $0.48\pm0.02$         & $4.2\pm0.3$           & $100/76$  & 0.03 \\
8   & $7.1\pm0.9$           & $1.3\pm0.1$           & $47\pm21$             & $5.0\pm1.0$           & 
    $4.2\pm0.6$             & $-10.415\pm0.006$     & $1.50\pm0.04$         & $0.80\pm0.05$         &
    $-11.48\pm0.01$         & $0.62\pm0.09$         & $2.9\pm0.5$           & $109/86$  & 0.05 \\
10   & $4.7\pm0.8$          & $1.1\pm0.2$           & $2.2\pm1.3$           & $3.7\pm0.4$           & 
    $2.7\pm0.6$             & $-10.14\pm0.01$       & $1.64\pm0.05$         & $0.92\pm0.04$         &
    $-11.33\pm0.03$         & $0.57\pm0.03$         & $5.3\pm0.6$           & $71/76$   & 0.65 \\
11   & $4.7\pm0.3$          & $1.1\pm0.1$           & $0.9\pm0.7$           & $3.2\pm0.4$           & 
    $1.6\pm0.3$             & $-10.136\pm0.005$     & $1.54\pm0.03$         & $0.89\pm0.05$         &
    $-11.36\pm0.03$         & $0.57\pm0.03$         & $5.3\pm0.4$           & $96/86$   &0.21 \\
12   & $6.3\pm0.5$          & $1.1\pm0.1$           & $6.1\pm2.2$           & $3.4\pm0.1$           & 
    $8.4\pm0.5$             & $-9.996\pm0.005$      & $1.62\pm0.03$         & $0.80\pm0.02$         &
    $-11.38\pm0.04$         & $0.67\pm0.09$         & $3.8\pm0.6$           & $64/77$   & 0.85 \\
13   & $5.6\pm1.8$          & $1.2\pm0.5$           & $4.5\pm3.3$           & $4.0\pm0.5$           & 
    $9.8\pm1.0$             & $-10.364\pm0.008$     & $1.37\pm0.07$         & $0.81\pm0.06$         &
    $-11.43\pm0.02$         & $0.56\pm0.03$         & $4.6\pm0.5$           & $81/75$   & 0.29 \\
15   & $9.5\pm0.7$          & $1.5\pm0.2$           & $1.1\pm0.5$           & $3.6\pm0.4$           & 
    $2.1\pm0.5$             & $-10.09\pm0.01$       & $1.55\pm0.05$         & $0.86\pm0.05$         &
    $-11.36\pm0.03$         & $1.2\pm0.2$           & $2.3\pm0.3$           & $67/79$   & 0.83 \\
16   & $4.3\pm1.3$          & $0.6\pm0.2$           & $9.4\pm3.0$           & $4.7\pm0.6$           & 
    $2.8\pm0.8$             & $-10.376\pm0.007$     & $1.46\pm0.03$         & $0.89\pm0.01$         &
    $-11.48\pm0.02$         & $0.52\pm0.01$         & $4.7\pm0.1$           & $102/77$  & 0.027 \\
17   & $2.6\pm1.6$          & $0.6\pm0.7$           & $32\pm10$             & $4.3\pm0.1$           & 
    $26.6\pm0.7$            & $-10.275\pm0.010$     & $1.52\pm0.06$         & $0.95\pm0.04$         &
    $-11.45\pm0.01$         & $0.46\pm0.03$         & $4.7\pm0.5$           & $68/72$   & 0.60 \\
18   & $6.0\pm0.8$          & $1.1\pm0.2$           & $13\pm7$              & $4.3\pm0.4$           & 
    $27.1\pm1.0$            & $-10.405\pm0.005$     & $1.47\pm0.04$         & $0.88\pm0.01$         &
    $-11.51\pm0.01$         & $0.53\pm0.01$         & $4.3\pm0.3$           & $71/78$   & 0.70 \\
19   & $2.1\pm0.6$          & $-0.6\pm0.7$          & $19\pm10$             & $5.0\pm0.7$           & 
    $33.4\pm1.0$            & $-10.32\pm0.01$       & $1.44\pm0.07$         & $0.92\pm0.01$         &
    $-11.45\pm0.02$         & $0.52\pm0.01$         & $4.8\pm0.1$           & $72/71$   & 0.45 \\
20   & $3.3\pm0.5$          & $-0.1\pm0.3$          & $13\pm4$              & $4.8\pm0.5$           & 
    $25.8\pm0.7$            & $-10.177\pm0.008$     & $1.59\pm0.05$         & $0.94\pm0.01$         &
    $-11.47\pm0.02$         & $0.56\pm0.01$         & $4.3\pm0.3$           & $101/76$  & 0.027 \\
21   & $1.7\pm0.3$          & $-0.8\pm0.3$          & $11\pm6$              & $4.2\pm0.1$           & 
    $26.7\pm0.4$            & $-10.156\pm0.004$     & $1.57\pm0.04$         & $0.95\pm0.01$         &
    $-11.44\pm0.01$         & $0.36\pm0.02$         & $6.6\pm0.6$           & $69/73$   & 0.62 \\
22   & $3.7\pm1.0$          & $0.5\pm0.3$           & $12.0\pm5.0$          & $4.1\pm0.1$           & 
    $28.6\pm0.7$            & $-9.992\pm0.007$      & $1.63\pm0.04$         & $0.95\pm0.01$         &
    $-11.40\pm0.01$         & $0.49\pm0.01$         & $5.3\pm0.3$           & $79/76$   & 0.39 \\
23   & $5.8\pm0.5$          & $1.2\pm0.1$           & $6.3\pm1.0$           & $3.5\pm0.1$           & 
    $27.7\pm0.4$            & $-9.957\pm0.005$      & $1.60\pm0.03$         & $0.93\pm0.01$         &
    $-11.40\pm0.02$         & $0.57\pm0.03$         & $4.3\pm0.4$           & $140/85$  & $10^{-4}$ \\
24   & $5.5\pm0.7$          & $1.2\pm0.1$           & $4.1\pm1.8$           & $3.7\pm0.2$           & 
    $28.1\pm0.4$            & $-10.062\pm0.006$     & $1.62\pm0.03$         & $0.93\pm0.01$         &
    $-11.39\pm0.02$         & $0.57\pm0.04$         & $4.4\pm0.4$           & $97/81$   & 0.11 \\
\enddata
\end{deluxetable}
\end{longrotatetable}
\newpage





\bibliography{main}

\listofchanges

\end{document}